%% file: main.tex
\documentclass{article}

\PassOptionsToPackage{numbers, compress}{natbib}

\usepackage[final]{neurips_2025}

\usepackage{enumitem}
\usepackage[utf8]{inputenc} 
\usepackage[T1]{fontenc}    
\usepackage{hyperref}       

\usepackage{url}            
\usepackage{booktabs}       
\usepackage{amsfonts}       
\usepackage{nicefrac}       
\usepackage{microtype}      
\usepackage{xcolor}         
\usepackage{graphicx}
\usepackage{algorithm}
\usepackage{algorithmic}
\usepackage{multicol}
\usepackage{multirow} 
\usepackage{pifont}
\usepackage{amsmath}    
\usepackage{amssymb}    
\usepackage[capitalize]{cleveref}
\usepackage{array}
\usepackage{subcaption} 
\usepackage{float}
\title{Sequential Attention-based Sampling for Histopathological Analysis}

\usepackage{color}
\usepackage{soul}
\usepackage{makecell}
\author{%
  Tarun Gogisetty \\
  Indian Institute of Science, Bangalore, India \\
  \texttt{tarung@iisc.ac.in}
  \And
  Naman Malpani \\
  Indian Institute of Science, Bangalore, India \\
  \texttt{namanmalpani@iisc.ac.in}
  \And
  Gugan Thoppe \\
  Indian Institute of Science, Bangalore, India \\
  \texttt{gthoppe@iisc.ac.in}
  \And
  Sridharan Devarajan\\
  Indian Institute of Science, Bangalore, India \\
  \texttt{sridhar@iisc.ac.in}
}

\begin{document}
\newcommand{\noref}[1]{\ref*{#1}}

\maketitle
\input{sec/0_abstract}    
\input{sec/1_intro}
\input{sec/2_related_work}

\input{sec/3_methods.tex}
\input{sec/4_experiments.tex}

{
    \small
    \bibliographystyle{plainnat}
    \bibliography{main}
}


\newpage
\input{sec/6_appendix}

\end{document}

%% file: sec/0_abstract.tex
\vspace{-5mm}
\begin{abstract}

Deep neural networks are increasingly applied in automated histopathology. Yet, whole-slide images (WSIs) are often acquired at gigapixel sizes, rendering them computationally infeasible to analyze entirely at high resolution. Diagnostic labels are largely available only at the slide-level, because expert annotation of images at a finer (patch) level is both laborious and expensive. Moreover, regions with diagnostic information typically occupy only a small fraction of the WSI, making it inefficient to examine the entire slide at full resolution.
Here, we propose SASHA -- {\it S}equential {\it A}ttention-based {\it S}ampling for {\it H}istopathological {\it A}nalysis -- a deep reinforcement learning approach for efficient analysis of histopathological images. 
First, SASHA learns informative features with a lightweight hierarchical, attention-based multiple instance learning (MIL) model. 
Second, SASHA samples intelligently and zooms selectively into a small fraction (10-20\%) of high-resolution patches to achieve reliable diagnoses.
We show that SASHA matches state-of-the-art methods that analyze the WSI fully at high resolution, albeit at a fraction of their computational and memory costs. In addition, it significantly outperforms competing, sparse sampling methods. 
We propose SASHA as an intelligent sampling model for medical imaging challenges that involve automated diagnosis with exceptionally large images containing sparsely informative features\footnote{Model implementation is available at: \url{https://github.com/coglabiisc/SASHA}}.

\end{abstract}

%% file: sec/1_intro.tex
\vspace{-5mm}
\section{Introduction}
\label{sec:intro}
\vspace{-2mm}
Deep learning models have emerged as a major frontier in automated medical imaging diagnosis. For example, such models accurately identify tumorous tissue in whole-slide histopathological images (WSI) \citep{hou2016patch, gecer2018dcnc, hameed2020breast, li2021multi}. Yet, hardware and memory bottlenecks render it impractical to analyze WSIs acquired at gigapixel sizes. Moreover, analyzing WSI-s at high resolution is inefficient because regions with diagnostic information (e.g., tumor cells) make up only a small fraction of the entire image, and are not clearly identifiable at low resolutions. As a result, methods that predict slide-level labels from low-resolution WSIs suffer from performance bottlenecks \cite{ruan2021btl1, jenkinson2024btl2}. By contrast, methods that incorporate fine-grained ``patch-level'' information perform better \cite{roy2019patch, wang2021tumor}. Yet, these latter methods require significant resources and expert annotations of WSIs at the patch level.

The Multiple Instance Learning (MIL) framework \cite{das2018multiple}, addresses these limitations elegantly. In MIL, the WSI is divided into non-overlapping patches, with each patch being encoded into a high-dimensional feature vector. The feature vectors are aggregated using simple methods, like max- or mean-pooling, and this ``bag'' of features is slide-level labels. Furthermore, diagnostic regions can be identified with weighted aggregation methods, like Attention-Based deep Multiple Instance Learning (ABMIL)~\cite{ilse2018abmil}, which assigns patch-level attention scores, based on their relevance.

\begin{figure*}
    \centering

    \resizebox{\textwidth}{!}{
        \includegraphics{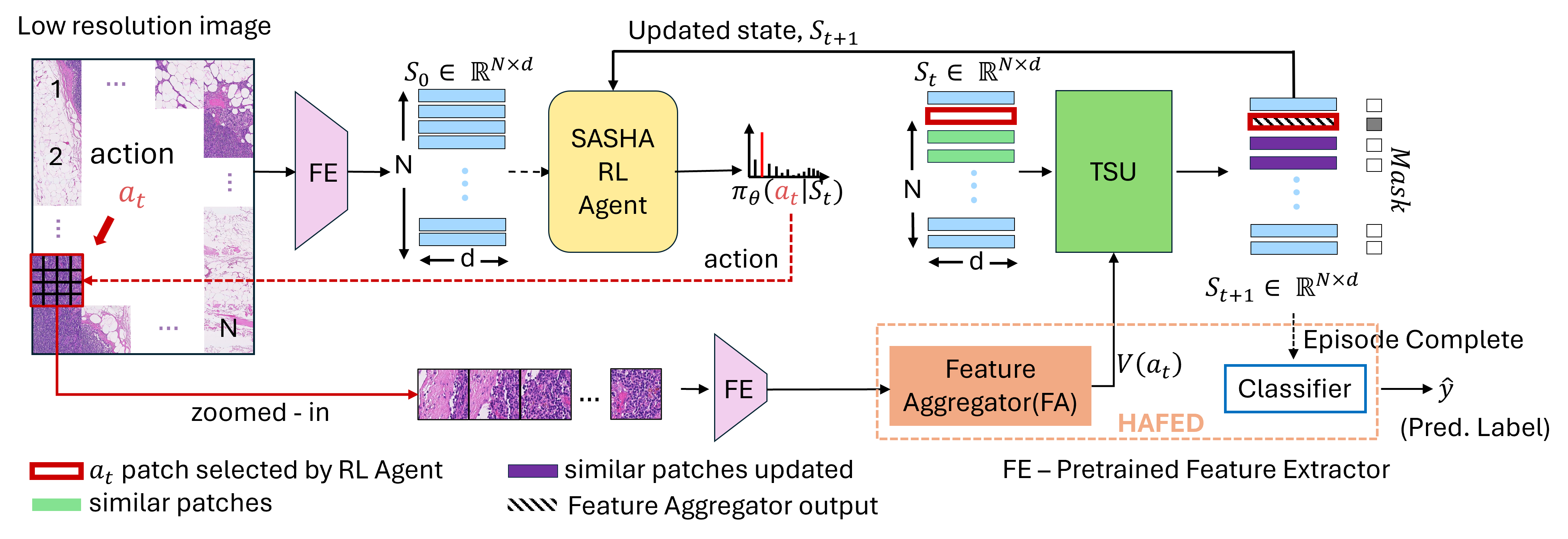} 
    }
    
    \caption{SASHA: An attention-augmented deep RL model. Starting with low-resolution WSI features \( S_0 \in \mathbb{R}^{N \times d} \),  at each time step \( t \), the deep RL agent selects a patch $a_t$ (red outline) to zoom in at high-resolution, based on policy $\pi(a_t|S_t)$. High-resolution features \( V(a_t) \in \mathbb{R}^{d} \) are extracted and aggregated using a Heirarchical Attention-based Feature Distiller (HAFED, \cref{sec:fac}). The states $S_t$ of patches with features similar to $a_t$ are updated with a Targeted State Updater (TSU, \cref{sec:ssu}). At the end of each episode, the classifier (\cref{sec:fac}) predicts the presence or type of cancer.}

    \label{fig:semantic_dig}


    \vspace{-5mm}
\end{figure*}



Nonetheless, a major limitation of MIL-based approaches is the need to process most WSI patches at high resolution. 
Because tumors typically occupy only a small portion of the slide, feature extraction for the majority of non-tumor patches becomes computationally inefficient, increasing inference time and memory usage. 
To address this, Zhao et al.~\cite{zhao2022rlogist} proposed a deep reinforcement learning (RL)-based approach -- RLogist, that first examines WSIs at low resolution and then selectively zooms into diagnostically relevant high-resolution patches to predict slide-level labels. 

Although reportedly much faster ($\sim$4x) than conventional MIL approaches at inference time~\cite{zhao2022rlogist}, RLogist suffers from major drawbacks: 
i) Poor accuracy: Because RLogist samples only a fraction of the whole slide, it lags current sota models that sample the whole WSI at high resolution by $\sim$10-15$\%$; 
ii) Weakly diagnostic features: Image features are  
 extracted with a generic ResNet-50 \cite{he2016deep} model pretrained on ImageNet \cite{imagenet}, thereby rendering these features poorly diagnostic of pathology; 
iii) Training inefficiency: The RL agent training suffers from bottlenecks, 
such as convergence issues arising from the RL policy network being trained concurrently with the classification network.

Here, we address these limitations by combining RL with multi-attention MIL in a framework we call SASHA - {\bf S}equential {\bf A}ttention-based {\bf S}ampling for {\bf H}istopathological {\bf A}nalysis (\cref{fig:semantic_dig}). Our study makes the following key contributions:
\begin{itemize}[leftmargin=*]  
    \itemsep0em 
    \item We design an attention-augmented deep RL agent that scans WSI patches at low resolution, and zooms into selected patches to extract meaningful, high-resolution features. 
    \item To improve diagnostic accuracy, we propose a Hierarchical Attention-based FEature Distiller (HAFED), which uses multiple attention heads to generate label-informed MIL features. 
    \item For stable convergence of the RL policy, we train a slide-level label classifier as part of the HAFED model, and freeze the classifier during RL training.
    \item For faster training and inference, we exploit similarities among patch features to perform a concerted state update of correlated instances with a targeted state updater (TSU).
    
\end{itemize} 
We show that SASHA achieves state-of-the-art performance on two cancer benchmarks \citep{bejnordi2017diagnostic, tcga2014cancer}, comparable to models that analyze the entire WSI at high resolution, while requiring only $\sim 6 \%$ of their memory for WSI representation and with upto 8$\times$ faster inference times.

%% file: sec/2_related_work.tex
\vspace{-2mm}
\section{Related Work}
\label{sec:related work}
\vspace{-2mm}
Deep learning approaches for automated histopathology have a rich and detailed history (reviewed in \cite{srinidhi2021deep, Qu_2022_DL_histo}). Here, we review specifically those methods that inspire key aspects of our solution and, therefore, form benchmarks for performance comparisons.

\noindent{\bf Multiple Instance Learning (MIL)}. MIL is a weakly supervised learning paradigm widely applied in  WSI analysis. Unlike fully supervised learning, where each instance is associated with a label, MIL involves dividing the slide into a ``bag'' of instances (patches); the model is then trained with bag-level labels. MIL is, therefore, suitable for classification problems where slide-level, but not patch-level, labels are available. Early MIL approaches relied on instance selection to identify the most relevant patches within a bag (\cite{GADERMAYR2024_MIL_review}). Recent models, such as attention-based MIL (ABMIL) \cite{ilse2018abmil}, assign importance weights to patches, allowing the model to focus on diagnostically relevant regions. 

{\bf Advances in attention-based MIL.} 
A key limitation of these approaches is that each patch is treated independently of the others, even though many may share strong feature correlations. Recent models exploit correlations among instance representations using a self-attention mechanism (e.g., TransMIL~\cite{shao2021transmil}). Other approaches cluster instances based on their attention scores~\cite{lu2021clam}.  
Yet, these attention mechanisms may capture only some types of diagnostic features, leading to overfitting and lack of generalization. To overcome this, Attention-Challenging Multiple Instance Learning (ACMIL)~\cite{zhang2023acmil} employs multiple attention blocks, each focusing on distinctive diagnostic features. Inspired by these approaches, our model incorporates multiple attention heads for diagnostically relevant feature extraction and a correlation-based state update rule for efficient feature updates. Moreover, earlier approaches largely process every patch at high resolution, which is computationally inefficient. ZoomMIL \cite{zoom_mil} addresses this by recursively sampling a fixed number of patches per resolution based on top-$k$ attention scores. However, its sampling strategy depends on predefined attention criteria, which motivates the use of adaptive and context-aware patch selection methods.

{\bf Deep Reinforcement Learning (DRL) in medical imaging}. Conventional RL learns a policy for actions that maximizes the expected reward in novel, unstructured environments. DRL augments RL with the representational power of deep neural networks. The power of DRL was originally demonstrated in challenging strategy and video games (e.g., Deep-Q, AlphaGo), \cite{mnih2013playingatariDRL, silver2017mastering}. DRL is particularly useful in domains with limited annotations, such as automated medical diagnosis, and has been successfully applied to various medical imaging tasks, including localization of anatomical landmarks in medical scans \citep{ALANSARY2019156}, lesion detection \citep{Maicas2017lesion, stember2020lesion}, 3D image segmentation of MRI scans \citep{liao2019segmentation} and tumor classification in WSIs \citep{zhao2022rlogist, zheng2024cvpr}. In such tasks, a DRL agent learns an optimal search strategy by exploring different portions of the environment (here, image) and receiving sparse feedback in the form of classification accuracy, without the need for analyzing the entire image at high resolution.

{\bf DRL for histopathology}. 
Diagnosis of WSIs by human histopathologists involves viewing the slide -- at a low (``scanning'') resolution -- and zooming into suspected tumor regions -- at high resolution -- for reliable diagnosis. In this context, a recent approach, RLogist~\citep{zhao2022rlogist}, models the scanning process of histopathologists as a Markov Decision Process (MDP) using an RL agent. The agent learns an efficient observation strategy, selectively zooming into only a few high-resolution WSI patches before making a classification decision, thereby reducing computational overhead and inference time. Similarly, PAMIL \citep{zheng2024cvpr} addresses the shortcomings of instance-based clustering methods using an RL agent to form pseudo-bags (aggregates) of related instances from high-resolution WSIs. It models pseudo-bag formation as an MDP, forming new pseudo-bags based on past knowledge from the previously formed bags, informed by the label error arising from bag-level predictions. 






{\bf Feature extraction for histopathology}. Deep neural networks such as ResNet \citep{he2016deep} and Vision Transformers (ViT) \citep{doso2020vit}, pre-trained on ImageNet \citep{imagenet}, are commonly used for feature extraction in computer vision tasks. Their exposure to millions of natural images enables them to identify generic, relevant features, for specialized downstream tasks. Yet, pretraining on natural images is not ideal for medical image diagnosis.
\cite{REMEDIS}. For instance, histopathology images lack a canonical orientation, exhibit low color variation due to standardized staining protocols, and their diagnostic information varies with magnification. To address this, Kang et al.~\cite{kang2023ssl} introduced ResNet-50 and ViT-S models with self-supervised pretraining on large WSI datasets (e.g., TCGA, TULIP).

%% file: sec/3_methods.tex
\vspace{-3 mm}
\section{Model description}
\label{sec:method}
\vspace{-2 mm}
Our framework combines multi-attention multiple instance learning (MIL) with reinforcement learning (RL), to identify diagnostic patches in the WSI. We also employ a targeted update rule that exploits correlations among instances to efficiently update the state. Each step in the model -- and how it addresses shortcomings with existing methods -- is described subsequently.
\vspace{-2 mm}

\begin{figure*}[t]
    
    \centering
    
    \resizebox{\textwidth}{!}{
        \noindent\includegraphics{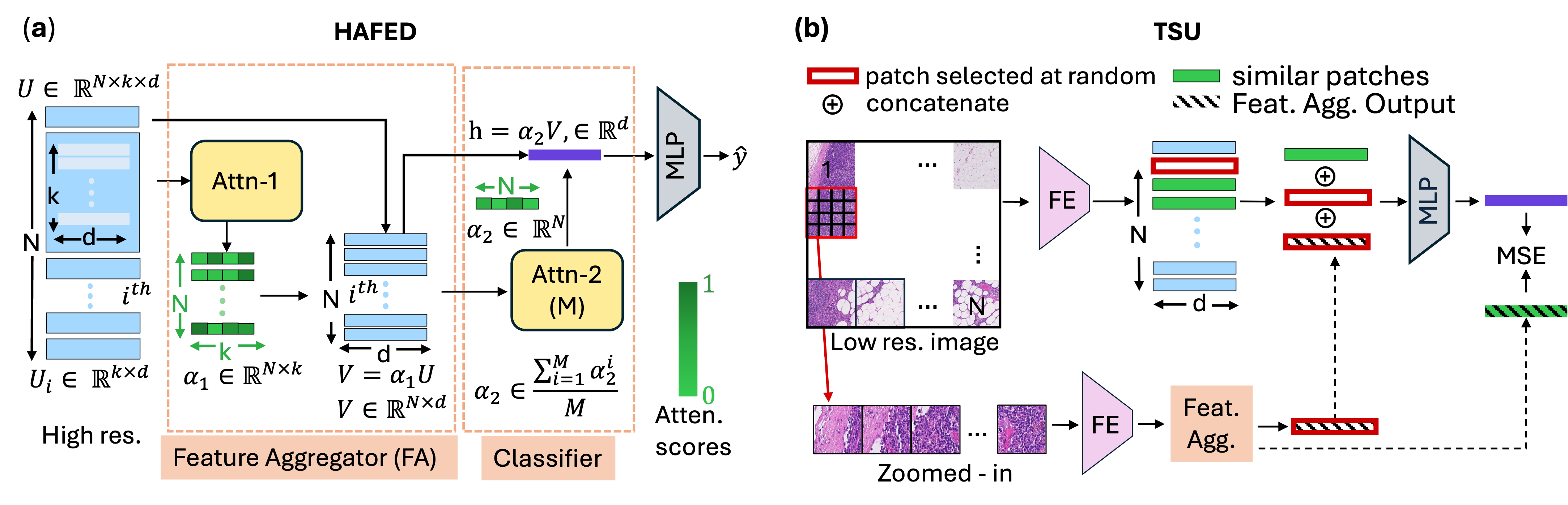} 
    }
    
    \caption{Components of SASHA. ({\bf a}) {HAFED}. During training, features from each patch in each WSI are extracted and aggregated with a hierarchical, attention-based feature distiller (HAFED). High-resolution input \( U \in \mathbb{R}^{N \times k \times d} \), undergoes two levels of attentional filtering, first operating across the $k$ high resolution patches in each low resolution patch $\alpha_1 \in \mathbb{R}^{N \times k} $ and then operating across the $N$ low resolution patches $\alpha_2 \in \mathbb{R}^N$ to produce the final classifier prediction $\hat{Y}$. During inference, only the sampled patch's high-resolution features are analyzed. ({\bf b}) {TSU}. Following this the state of all patches whose features are correlated with the sampled patch are concurrently updated, with a targeted state updater (TSU) (see sections \ref{sec:fac} and \ref{sec:ssu} for details).
    }

    \label{fig:hafed_tsu}
    \vspace{-5mm}
\end{figure*}

\vspace{-1mm}
\subsection{Problem Statement and Overview}
\label{sec:problem_statement}
\vspace{-2mm}
For our task, the model seeks to predict binary target labels \( Y \in \{0, 1\} \) for a given Whole Slide Image (WSI) \( X \). The latter is a 3-channel RGB image that is available in the dataset at multiple different levels of magnification, whereas the former either indicates the presence (versus absence) of cancer or indicates one of two subtypes. We notate the size (in pixels) of the image at magnification level $m \in \{1,2,3\}$ as $G^m$. The goal of our approach is to scan the image at the lowest resolution $G^3$, which we call the ``scanning-level'' as in~\cite{zhao2022rlogist}, while zooming into and analyzing only select portions of the image at a higher resolution $G^1$. For this, each WSI at $G^3$ is divided into multiple non-overlapping patches, such that a given WSI \( X \) is represented as \( X = \{ x_1, \ldots, x_N \} \), where \( N \) is the total number of patches at low resolution $G^3$; note that $N$ can vary across different WSI images. Now, when zoomed in, these patches are magnified to a high-resolution; for this, we choose the highest resolution $G^1$ in all datasets tested. Therefore, each patch \( x_i \) at the low-resolution comprises of  \( k \) patches at the high-resolution; we denote these as \(\{ x_{i1}, \ldots, x_{ik} \} \). 

Given these definitions, the three main components of the algorithm are as follows (\cref{fig:semantic_dig}):

    {\bf Hierarchical Feature Distiller:} This hierarchical attention-based network extracts diagnostic features from the $k$ high-resolution patches for each low-resolution patch and creates a $d$ dimensional feature representation for it. The output of this network is used to update WSI ``state'' and to train a downstream WSI label classifier; see \cref{sec:fac} for details.

    {\bf Targeted State Updater:} This network maintains a state of the entire WSI, comprising a $d$ dimensional feature representation for each low-resolution patch. As each patch is zoomed in to high-resolution, the state vector is updated, but only for instances whose features are correlated with those of the sampled patch; see \cref{sec:ssu} for details.

    {\bf Deep RL Agent:} This agent comprises a policy network and a value network. Based on the current state, the former network produces a probability distribution over a set of actions from which the agent selects the next patch to zoom in, whereas the latter network is used to compute a generalized advantage estimate for each state-action combination. Both networks are optimized using the Proximal Policy Optimization (PPO) algorithm; see \cref{sec: sasha} for details.  


\noindent{\bf Preprocessing and Feature Extraction}. Prior to training and testing, we preprocess the WSI and extract features using a pre-trained model, as follows: To isolate histopathological tissue from the glass slide background on the WSI we apply established segmentation methods \citep{lu2021clam} at low resolution. The extracted tissue is then divided into non-overlapping patches of size \( (256 \times 256 \times 3) \), again at both resolutions; these operations were performed using the code provided as part of the CLAM \citep{lu2021clam} GitHub repository. For the first level of feature extraction, we utilize a pre-trained ViT-based encoder \citep{doso2020vit}, as described in the Related Work (\cref{sec:related work}). Each input patch is encoded into a feature embedding vector of dimension \( {d} \). We denote the input feature representation of a WSI at low-resolution as \( Z \in \mathbb{R}^{N \times d} \), and at high-resolution as \( U \in \mathbb{R}^{N \times k \times d} \).

\vspace{-3 mm}
\subsection{Hierarchical Attention-based Feature Distiller (HAFED)}
\label{sec:fac}
\vspace{-2 mm}
A key first step with diagnosis of large WSIs is to learn informative, yet parsimonious, feature representations for the high-resolution patches. Such representations are critical for the RL agent to update WSI state, as it makes decisions about which patch to visit next, as well as for the final classification. Recent advances in attention-based MIL models compute such informative representations for high-resolution whole-slide images (WSIs) \cite{ilse2018abmil}. However, in an RL setting, the state representation for the WSI at each time step comprises a mix of information both at low and high-resolutions, depending on the specific patches zoomed in (see next). It is, therefore, desirable for both the low-resolution and high-resolution feature representations to have matched dimensionality; such a design choice considerably simplifies downstream computations (e.g., classification). To achieve this, we propose a Hierarchical Attention-based FEature Distiller (HAFED)  (\cref{fig:hafed_tsu}a). 

HAFED is a two-stage attention-based model designed to operate directly on high-resolution feature patches while producing representations compatible with low-resolution patches. In the first stage (Fig. \ref{fig:hafed_tsu}a, Feature Aggregator), the model takes as input high-resolution features \(U \in \mathbb{R}^{N \times k \times d}\), and learns to estimate attention scores for each of the \(k\) sub-patches, reflecting the latter's relative importance in each of the $N$ low resolution patches; a stochastic instance masking strategy is applied to avoid over-fitting \cite{zhang2023acmil}. A weighted aggregation of the sub-patches based on these attention scores produces a compressed feature representation \(V \in \mathbb{R}^{N \times d}\), thereby reducing the dimensionality of the high-resolution features and aligning them with the dimensionality of the low-resolution features \(Z \in \mathbb{R}^{N \times d}\). In the second stage (Fig. \ref{fig:hafed_tsu}a, Classifier), a second attention mechanism operates across the \(N\) patch-level features in \(V\) to identify diagnostically important patches at the slide level. A weighted aggregation of the \(N\) features, guided by the second-stage attention scores, yields a final slide-level embedding \(h\ \in \mathbb{R}^{d}\) that summarizes the entire WSI. This $d$-dimensional embedding is then used for downstream WSI classification. We employed multiple ($M$) attention branches, and the model is trained with the sum of three losses \cite{zhang2023acmil}: i) a similarity loss to learn distinct attention weight patterns in each branch, ii) the label loss associated with each branch, and iii) overall label loss.


During training, both stages of the model are trained end-to-end by visiting all of the $N$ patches at high resolution. Thus, attention weights, and feature representations, in both the first and second stage of the model are learned in a ``label-aware'' manner using the ACMIL loss (classifier and dissimilarity losses). During inference, a single high-resolution patch $a_t$ is visited at each timestep $t$ and its compressed feature representation \(V(a_t) \in \mathbb{R}^{d}\) is used to update the slide state. While previous approaches \citep{zhao2022rlogist} employ generic, label-agnostic feature extractors HAFED identifies important regions at both the sub-patch (first-stage) and patch (second-stage) levels, which enables efficient learning of diagnostically-relevant and parsimonious feature representations from high-resolution WSIs.


    
\vspace{-2 mm}
\subsection{Targeted State Updater (TSU)}
\label{sec:ssu}
\vspace{-2 mm}

In our deep RL model, each WSI is represented by a ``state'' \(S_t \in \mathbb{R}^{N \times d}\) at each timestep \(t\) in an episode. The RL agent makes sequential decisions of which patch to visit based on \(S_t\). We describe, here, how this state is computed and updated over successive time steps: Scanning begins with the low-resolution feature representation \(Z\), which serves as the initial state \(S_0\). While \(Z\) provides a coarse global view of the WSI, it often lacks the fine-grained diagnostic details necessary for accurate classification. Obtaining high-resolution features for all patches is computationally expensive; thus, at each step \(t\), we selectively zoom into a single patch at high resolution to gather additional information. The selected patch, denoted as \(a_t\), yields \(k\) high-resolution sub-patches, which are processed through the first attention model of HAFED. A weighted aggregation of the sub-patches produces a refined high-resolution feature vector \(V(a_t)\ \in \mathbb{R}^{d}\) that captures detailed local information.

The challenge then is how to effectively integrate this newly acquired high-resolution information into the overall state representation \(S_t\) without corrupting  previously acquired information. One possible approach (e.g., \citet{zhao2022rlogist}) involves concatenating \(V(a_t)\) with every patch in \(S_t\), training an MLP  (multi-layer perceptron) with this augmented state as input to produce an updated global state representation \(S_{t+1}\) that gradually converges toward the feature representation obtained from all high-resolution patches. We argue that such a global state update is far from ideal. Because only one high resolution patch is observed in each time step, updating instances globally risks corrupting the state, especially when instances unrelated to the sampled patch are concurrently updated. Moreover, information from the sampled patch is likely to be relevant, primarily, for updating instances that share features similar to it \citep{shao2021transmil, lu2021clam}.

To exploit these correlations, we propose a targeted, similarity-based state update (TSU) model (\cref{fig:hafed_tsu}b). For the selected patch \(a_t\), we compute the cosine similarity between \(S_t(a_t)\) and all other patch representations in \(S_t\). We then define a set \(C\) containing the indices of patches whose cosine similarity with \(S_t(a_t)\) exceeds a threshold (\(\tau\), a hyperparameter), excluding \(a_t\) itself. Only the patches in \(C\) are updated: \(V(a_t)\) is concatenated with the representations of patches in \(C\) and passed through an MLP trained to minimize the MSE loss against the corresponding entries in \(V(C)\). Additionally, \(S_t(a_t)\) is directly replaced with \(V(a_t)\) because its exact high-resolution representation is available. In the next time step, the patch \(a_t\) is masked out, and future actions are sampled from the remaining unmasked patches; this strategy is applied during both the training and evaluation stages. With ablation studies (\Cref{table:ablation}), we show that this selective update strategy outperforms naive approaches.


\begin{algorithm}[t]
\caption{Agent-Environment Loop at Time Step \( t \)}
\label{algo:agent_env_step}
\begin{multicols}{2}

\begin{algorithmic}

        \STATE \textbf{Input:} Current state $S_t \in \mathbb{R}^{N \times d}$; Set $I$ of sampled patch indices, upto time $t$ 
        \STATE \textbf{Parameters:} Policy network $\pi_{\theta_t}$; HAFED feature aggregator model $f_H(.; \theta_H)$ and classifier $f_C(.; \theta_C)$; TSU state update model $f_{S} (.; \theta_S)$
        \STATE \textbf{Output:} Updated state $S_{t+1}$ and action $a_{t+1}$ at time $t+1$
    
\end{algorithmic}
\begin{algorithmic}[1]

        \STATE $a_t \sim \pi_{\theta_t}(. \mid S_t)$ \COMMENT{Sample patch index $a_t$ from the policy distribution}
        \STATE $I \gets I \cup \{a_t\}$ \COMMENT{Store selected patch index}
        \IF{training}
            \STATE $\hat{y_t} \gets f_C(S_t; \theta_C)$ \COMMENT{predict WSI label with HAFED classifier}
            \STATE Update $\pi_{\theta}$ with a PPO algorithm; intermediate reward $r_t= - CE(y, \hat{y_t})$
        \ENDIF
        \STATE $V(a_t) \gets f_H(U_{a_t}; \theta_H)$ \COMMENT{extract features $V(a_t) \in \mathbb{R}^{d}$ from high-resolution zoomed in features, $U_{a_t} \mathbb{R}^{k \times d}$ of the sampled patch, with the HAFED model}
        \STATE $S_{t+1}(a_t) \gets V(a_t)$ \COMMENT{update the sampled patch with its high-resolution feature representation}
        \STATE $C \gets \{i: \cos \angle(S_t(i), S_t(a_t))\geq \tau\}$ \COMMENT{identify set of patch indices whose cosine similarity with $S_t(a_t)$ exceeds a threshold} 
        \FOR{each patch index $i$ in $C \setminus I$}
            \STATE $S_{t+1}(i) \gets f_{S}([S_t(i), S_t(a_t), V(a_t)]; \theta_S)$ \COMMENT{targeted update of similar patches with the TSU model}
        \ENDFOR
        \RETURN $S_{t+1}, a_{t+1}$
  
\end{algorithmic}
\end{multicols}
\end{algorithm}

\vspace{-2mm}
\subsection{Sequential Attention-based Sampling for Histopathological Analysis}
\label{sec: sasha}



\vspace{-2mm}
We train a reinforcement learning (RL) agent to select which patches to zoom into, leveraging feature distillation with the HAFED model and state update with the TSU model. We formulate the patch selection process as a sequential decision-making problem, where an agent interacts with a slide-specific environment over multiple steps (\cref{algo:agent_env_step}). During each training episode, at each timestep \(t\), the agent observes the WSI state \(S_t\), which is initialized with the low-resolution feature \(S_0 = Z\). Based on \(S_t\), the policy network  \(\pi_{\theta}\) produces a probability distribution over a set of actions from a discrete action space \(\{1, 2, \ldots, N\}\). The agent samples an action \(a_t\), corresponding to the next low-resolution patch to zoom in, according to the current policy \(\pi_{\theta_t}(\cdot \mid S_t)\) with a stochastic sampling rule. The zoomed-in patch is passed through HAFED to obtain its high-resolution representation \(V(a_t)\). This representation is then used by the TSU to selectively refine the current state \(S_t\), by modifying only those patches that are similar to \(a_t\) in the feature space. This produces an updated state \(S_{t+1}\). The classifier trained together with HAFED then uses the updated state \(S_{t+1}\) to produce a slide-level prediction \(\hat{y}_t\). This prediction is compared against the ground truth label \(y\) to compute a time-step-wise reward, based on a negative cross-entropy (CE) loss. This feedback is used to update the policy network using Proximal Policy Optimization (PPO) \citep{schulman2017ppo} algorithm. We employed Generalized Advantage Estimation (GAE) in computing the advantage function \citep{gae}, whose output was trained to predict the reward until the end of the episode. Both the policy and value network were modeled on standard architectures \cite{ilse2018abmil} (details in Appendix \ref*{appendix:actor_critic}). The same steps were followed during evaluation, except that the action with the highest probability based on the policy network's output was chosen deterministically. 

%% file: sec/4_experiments.tex
\vspace{-2mm}
\section{Experiments}
\vspace{-3mm}
\subsection{Dataset and implementation details}
\label{sec:dataset}
\vspace{-2mm}
We evaluate the effectiveness of SASHA with two publicly available benchmark datasets CAMELYON16 breast cancer \citep{bejnordi2017diagnostic} and TCGA-NSCLC lung cancer dataset \citep{tcga2014cancer}. While the CAMELYON16 labels correspond to normal versus pathological WSIs, TCGA-NSCLC labels correspond to distinguishing two different types of cancers, adenocarcinoma versus squamous cell carcinoma. We followed standardized train-test splits for both datasets. In addition, we also evaluate SASHA on multi-class classification tasks using the BRACS and Camelyon+ datasets. Further details on the datasets, hyperparameter selection, implementation, and training are in Appendices \ref*{appendix:dataset} and \ref*{appendix:implen_and_hy-params}.







\begin{table}[t] 
\centering

\resizebox{\textwidth}{!}{

\begin{tabular}{llcccccc}
\toprule
\multirow{2}{*}{} & \multirow{2}{*}{Method} & \multicolumn{3}{c}{CAMELYON16} & \multicolumn{3}{c}{TCGA - NSCLC} \\
& & Accuracy & AUC & F1 & Accuracy & AUC & F1 \\
\midrule

\multirow{7}{*}{\rotatebox[origin=c]{90}{Sample 100\%}} 
& Max Pooling       & 0.933 $\pm$0.007 & 0.971$\pm$0.001 & 0.911$\pm$0.008 & 0.881$\pm$0.010 & 0.952$\pm$0.008 & 0.882$\pm$0.010\\
& ABMIL             & 0.947$\pm$0.008 & 0.958$\pm$0.016 & 0.928$\pm$0.011 & 0.883$\pm$0.045 & 0.947$\pm$0.021 & 0.884$\pm$0.045\\
& CLAM\_SB          & 0.949$\pm$0.012 & 0.976$\pm$0.008 & 0.931$\pm$0.015 & 0.905$\pm$0.019 & 0.961$\pm$0.009 & 0.907$\pm$0.018\\
& TransMIL          & 0.936$\pm$0.006 & 0.953$\pm$0.008 & 0.912$\pm$0.008 & 0.898$\pm$0.015 & 0.964$\pm$0.004 & 0.899$\pm$0.016\\
& DTFD              & 0.944$\pm$0.014 & 0.977$\pm$0.008 & 0.944$\pm$0.014 & 0.911$\pm$0.012 & 0.962$\pm$0.010 & 0.911$\pm$0.022\\
& ACMIL             & 0.941$\pm$0.015 & 0.970$\pm$0.011 & 0.924$\pm$0.017 & 0.906$\pm$0.025 & 0.959$\pm$0.006 & 0.907$\pm$0.026\\
& {\bf HAFED} (ours)            & \underline{0.963}$\pm$0.008 & \underline{0.980}$\pm$0.003 & \underline{0.951}$\pm$0.011 & \underline{0.923}$\pm$0.011 & \underline{0.966}$\pm$0.015 & \underline{0.925}$\pm$0.010 \\

\midrule
\multirow{1}{*}{\rotatebox[origin=c]{90}{k}}
& Zoom-MIL (k=300)        & 0.780 $\pm$0.017 & 0.700 $\pm$0.029 &
0.738 $\pm$0.015 & 0.908 $\pm$0.004 &0.967 $\pm$0.004   &
0.908 $\pm$0.004 \\

\midrule

\multirow{2}{*}{\rotatebox[origin=c]{90}{\shortstack{10\%}}} 
& RLogist-0.1     & 
\makebox[1.85cm][l]{0.824} & 
\makebox[1.85cm][l]{0.829} & 
\makebox[1.85cm][l]{--} & 
\makebox[1.85cm][l]{0.828} & 
\makebox[1.85cm][l]{0.892} & 
\makebox[1.85cm][l]{--} \\
& Zoom-MIL (k=80)   & 0.753$\pm$0.02 & 0.743$\pm$0.007 & 0.743$\pm$0.012 & \textbf{0.907$\pm$0.002} & \textbf{0.969$\pm$0.012} & \textbf{0.907$\pm$0.002} \\
& \textbf{SASHA-0.1} (ours)   & \textbf{0.901$\pm$0.021} & \textbf{0.918$\pm$0.014} & \textbf{0.856$\pm$0.031} & {0.897$\pm$0.023} & 
{0.956$\pm$0.023} & 
{0.898$\pm$0.024} \\

\midrule

\multirow{2}{*}{\rotatebox[origin=c]{90}{\shortstack{20\%}}} 
& RLogist-0.2 & 
\makebox[1.85cm][l]{0.862} & 
\makebox[1.85cm][l]{0.879} & 
\makebox[1.85cm][l]{--} & 
\makebox[1.85cm][l]{0.839} & 
\makebox[1.85cm][l]{0.903} & 
\makebox[1.85cm][l]{--}  \\
& Zoom-MIL (k=160)   & 0.759$\pm$0.001 & 0.794$\pm$0.001 & 0.755$\pm$0.001 & 0.912$\pm$0.005 & 0.970$\pm$0.001 & 0.912$\pm$0.005 \\

& \textbf{SASHA-0.2} (ours)   & \textbf{0.953$\pm$0.017} & \textbf{0.979$\pm$0.008} & \textbf{0.937$\pm$0.024} & \textbf{0.912$\pm$0.010} & \textbf{0.963$\pm$0.014} & \textbf{0.914$\pm$0.011} \\
\bottomrule
\end{tabular}
}

\vspace{6pt}

\caption{Performance comparison on the CAMELYON16 and TCGA-NSCLC datasets. The top 7 rows reflect models that process all patches at high resolution (``Sample 100\%''). \underline{Underlined} values indicate the best performance among these methods. Rows 9-11 and 12-14 reflect models that selectively sample either ``10\%'' or ``20\%'' of the patches, respectively. \textbf{Bold} values indicate the best performance among these methods at the respective sampling fraction.}
\label{tab:classification_performance}
\vspace{-8mm}
\end{table}

\subsection{Baselines and Metrics}
\label{sec:baseline}
\vspace{-2mm}
We evaluate and compare the performance of our model against two types of baselines. First, we evaluate a range of different sota methods that analyze all patches in the WSI at high resolution, including MaxPooling, ABMIL \cite{ilse2018abmil}, CLAM  \cite{lu2021clam}, TransMIL \cite{shao2021transmil}, DTFD \cite{zhang2022dtfd}, and ACMIL \cite{zhang2023acmil}. We also include comparisons with our hierarchical attention-based (HAFED) model (\cref{sec:fac}) with the full-resolution WSI. Yet, these methods serve only as an upper bound on achievable performance.


For a fair comparison, we also evaluate models that sample only a fraction of the WSI. Specifically, we compare our approach with RLogist~\cite{zhao2022rlogist} when trained to sample exactly $10\%$ and $20\%$ of patches at high resolution, under identical experimental settings. We also compare with a non-RL-based Zoom-MIL~\cite{zoom_mil} model that samples exactly {\it k} = 80, 160, or 300 patches at low resolution (5x $\rightarrow$ 20x). The choice of {\it k} values aligns with the patch statistics of the dataset: WSIs contain  $\sim$800 low-resolution patches on average; thus, $k = 80$ and $k = 160$ correspond to $10\%$ and $20\%$ of the available patches, on average, providing a fair comparison with the RLogist-0.1/SASHA-0.1 and RLogist-0.2/SASHA-0.2 settings, respectively. $k = 300$ follows the default configuration used in the original Zoom-MIL paper~\cite{zoom_mil}. This setup allows us to systematically analyze the effect of varying observation budgets on model performance. In addition, we also compare against ACMIL and DTFD under the same observation budgets, with random sampling of high-resolution patches (\Cref{table:ablation}); details in Appendix \ref*{appendix:add_ablation}. Models were trained with the official codebase provided by the respective studies, except for RLogist. The RLogist codebase is insufficient for replication; therefore, we report the performance metrics {\it as is} from the original study \citep{zhao2022rlogist}.

Our performance metrics include Accuracy, AUC, and F1 scores for normal versus pathological tissue classification (CAMELYON16) or for classifying between the different tumor types (TCGA-NSCLC). In addition, we expect that selective sampling models, such as ours, would exhibit significantly faster inference times compared to models that examine all patches at high resolution; therefore, we also evaluate and compare inference times (\cref{fig:figure_3}c) . Finally, we also compute a metric of ``WSI compressibility''. We define this as a ratio of the original WSI image size, at full resolution, to the size the of WSI representation used by each model's classifier to make a final decision. This metric has important real-world implications in terms of efficient archival and storage of WSI images. We compare models of comparable complexity and performance (CLAM, TransMIL, ACMIL, DTFD, SASHA) in terms both WSI compressibility and overall model size.

\vspace{-2mm}
\subsection{Classification Performance, Inference Time and Compressibility}
\vspace{-2mm}
{\bf Classification Performance}. SASHA consistently outperforms the sota competing deep RL model -- RLogist -- by a large margin, for both the 10\% and 20\% observation budgets (\cref{tab:classification_performance}, SASHA-0.1 and SASHA-0.2, vs RLogist-0.1 and RLogist-0.2, respectively). Improved classification performance was reflected in all 3 metrics quantified: Accuracy, AUC, and F1-score. Zoom-MIL performed comparably with SASHA-0.2 for the TCGA benchmark, but underperformed heavily for CAMELYON-16. Moreover, our HAFED model performed on par with the best sota models (e.g. ACMIL, DTFD) that analyze the full, high-resolution WSI (\cref{tab:classification_performance}, underlined). 
Interestingly, even with sampling 5x fewer patches, SASHA-0.2 nears ceiling performance achievable with HAFED, when all 100\% of patches are sampled at high resolution  (\cref{tab:classification_performance}, compare SASHA-0.2 with HAFED).

\begin{figure}[t]
    \centering
    \includegraphics[height=0.3\textheight]{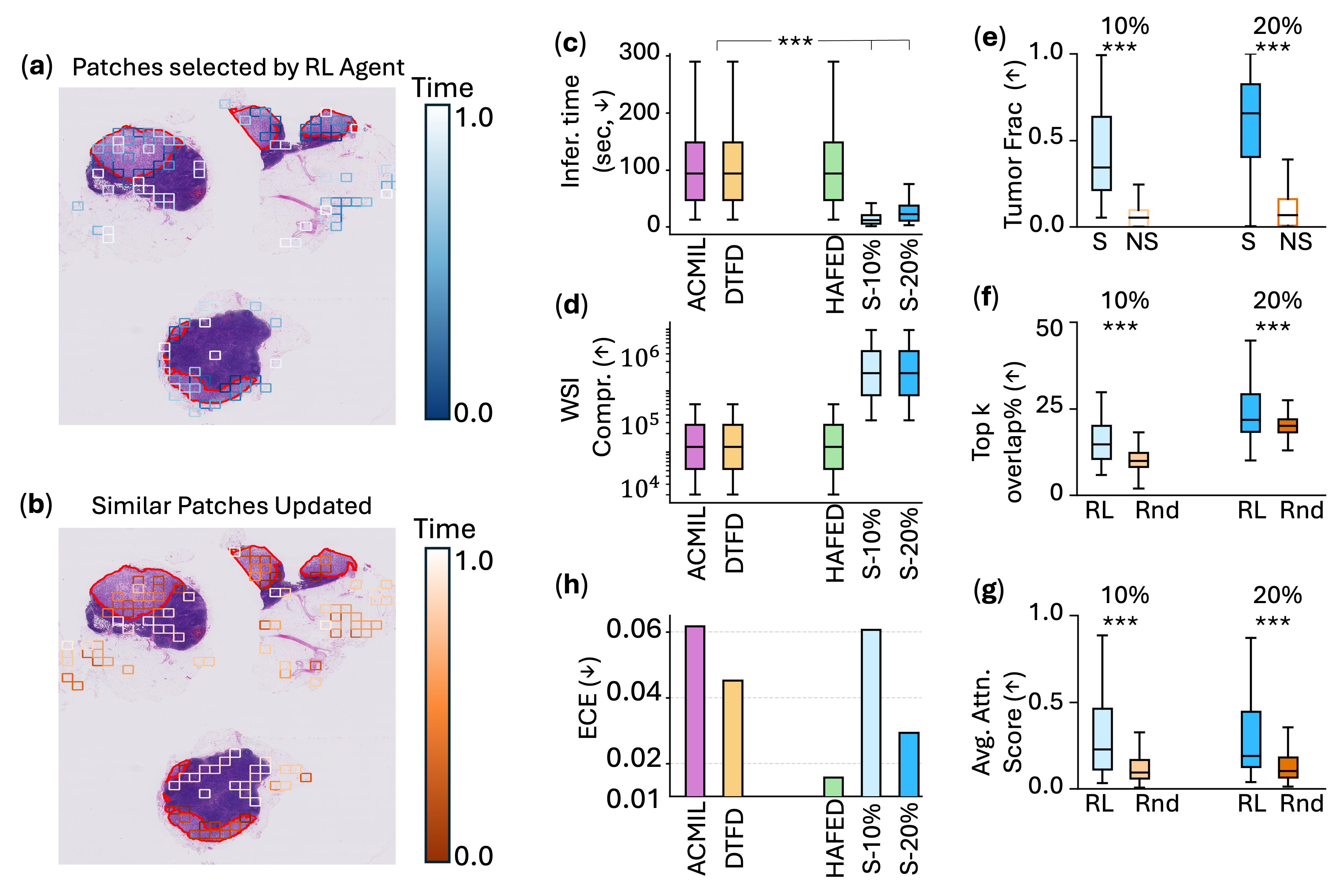}
    \caption{
    ({\bf a-b}) Patch selection and update strategy by SASHA's RL agent. Box color corresponds to time fraction in episode.
    ({\bf c-d, h}) Inference time, compressibility and expected calibration error (ECE) for SASHA and HAFED compared to other models. S-10\% and S-20\% refer to SASHA-0.1 and -0.2 models, respectively.
    ({\bf e}) Tumor fraction in patches sampled by SASHA's RL agent (S) versus non-sampled patches (NS).
    ({\bf f-g}) Top-k attention score overlap fraction and average attention score for patches sampled by SASHA's RL agent versus a random policy (Rnd). 
    In (e-g) left and right pairs of bars reflect 10\% and 20\% observation budgets, respectively.
    *** indicates $p < 0.001$. }
    
    \label{fig:figure_3}
    \vspace{-5mm}
\end{figure}

{\bf Inference time}. The average inference time per WSI for the CAMELYON16 dataset, including both pre-processing and feature extraction time, was $\sim$117 seconds, using HAFED (\cref{fig:figure_3}c). Other sota attention-based methods like ACMIL and DTFD also exhibited similar average inference times per WSI. The most expensive operation was feature extraction for each of the high resolution patches. By contrast, our model samples only a few patches at high-resolution, thereby leading to a substantial decrease in overall inference time, $\sim$14 seconds and $\sim$26 seconds per WSI for SASHA-0.1 and SASHA-0.2 -- an 8$\times$ and 4$\times$ speedup, respectively -- compared to the second fastest model (\cref{fig:figure_3}c). 

{\bf WSI compressibility}. Similarly, a comparison of the WSI compressibility revealed a significant ($>16\times$) improvement, compared to models that examine the entire slide at high resolution (\cref{fig:figure_3}d). The main reason for this saving is the fact that our HAFED model encodes features for both low- and high-resolution patches in a uniform $N \times d$ dimensional space; by contrast, other models encode high-resolution features in a $N \times k \times d$ dimensional space, where $k$ reflects the ratio of the zoomed-in to the scanning resolution size. 
The full set of results is shown in Appendix \ref*{appendix:comp_of}. These results highlight the efficiency of our method in balancing performance with computational and representational costs.

\vspace{-2mm}
\subsection{Ablation Study and Limited Sampling}
\vspace{-1mm}
We performed a systematic ablation study removing each component of the model in turn -- or replacing it with a naive variant -- and evaluating it with the CAMELYON16 dataset. These included: a) replacing the WSI-pretrained ViT with a ResNet50, used commonly for feature extraction \citep{zheng2024cvpr,zhao2022rlogist}, b) replacing ViT with the CONCH medical image-encoder\citep{lu2023conch}, pretrained on 1.17M image caption pairs, c) replacing multi-branch attention with single branch in the Classifer \citep{ilse2018abmil}, d) replacing the targeted state update policy with a global state update \citep{zhao2022rlogist}, e) removing the TSU and updating the state only of the sampled (local, high-resolution) patch, and f) and selecting an action based on a random policy (see ``Variant''s in \cref{table:ablation}). In every case, we observed sharp drops in accuracy, ranging from 7-45\%, relative to the baseline SASHA model; similar drops occurred in AUROC and F1 scores \cref{table:ablation}. The strongest impact occurred when changing the RL policy (\cref{table:ablation}, f). The next strongest impact occurred when changing the feature extractor to ResNet-50 (a), but this impact was mitigated when using the CONCH encoder (b). Intermediate levels of impact occurred when changing the classifier to single-branch attention (c) or altering the state update method (d-e). Additional ablation experiments are described in Appendix \ref*{appendix:add_ablation}. These results indicate that every component of our model was critical for achieving high performance, comparable to sota models.

Finally, we examined the effect of imposing a limited observation budget on other sota models, and compared the performance of SASHA under the same budget. Specifically, ACMIL and DTFD were trained and tested with a random selection of high-resolution patches under a 20\% observation budget for each WSI. In this case also we observed substantial drops in performance relative to the default SASHA model (\cref{table:ablation}, g-h). In other words, randomly sampling a limited fraction of high-resolution patches, even with the best models, failed to outperform SASHA.

\newcommand{\cmark}{\ding{51}} 
\newcommand{\xmark}{\ding{55}} 

\begin{table*}[t]
    \centering
        \resizebox{0.55\textheight}{!}{%
        \begin{tabular}{lccccccc}
        \toprule
        Variant & Feature & Classifier & TSU & RL Policy &  Accuracy & AUC & F1 \\
        \midrule
        SASHA Default                            & \cmark & \cmark & \cmark & \cmark &  0.964 & 0.980 & 0.953 \\
        (a) ResNet-50                         & *     & \cmark & \cmark & \cmark &  0.860 & 0.817 & 0.780 \\
        (b) CONCH                         & *      & \cmark & \cmark & \cmark &  0.930 & 0.950 & 0.905 \\
        (c) 1\textsuperscript{o} Att. Branch     & \cmark & *      & \cmark & \cmark &  0.899 & 0.964 & 0.851 \\
        (d) Global Update                      & \cmark & \cmark & *      & \cmark &  0.837 & 0.824 & 0.779 \\
        (e) Local Update                      & \cmark & \cmark & *      & \cmark &  0.860 & 0.901 & 0.800 \\
        (f) Random Policy                      & \cmark & \cmark & \cmark & *      &  0.516 & 0.550 & 0.553 \\ 
            \midrule

        (g) ACMIL-0.2 (Rnd.)                   & \cmark & \cmark & *       & *       & 0.838 & 0.857 & 0.773 \\  
        (h) DTFD-0.2 (Rnd.)                    & \cmark & \cmark & *       & *       & 0.862 & 0.878 & 0.862 \\ 
        \bottomrule
        \end{tabular}
        
        } 
        \caption{Ablation study on CAMELYON16 with the SASHA-0.2 setup using 20\% high-res patches. \cmark\ denotes the default, * the altered component.}
            \label{table:ablation}
\vspace{-5mm}

\end{table*}

    





\vspace{-2mm}
\subsection{Patch selection strategy: Explainability} 
\vspace{-2mm}
\label{sec:patch_sel_strategy}
When deploying deep RL models in medical settings, it is essential that they learn not only an effective, but also an explainable action policy. For this, we explore whether SASHA's RL agent exhibits a meaningful patch selection strategy prior to classifying WSI images (\cref{fig:figure_3}a and \cref{fig:figure_3}b). We propose that such a strategy must satisfy at least two important requirements: i) In WSIs containing tumor tissue, the RL agent must preferentially sample patches with a large fraction of tumor tissue for high-resolution scans, and ii) an effective RL policy should preferentially sample patches afforded a high attention score by sota models that examine the entire WSI at high resolution. We evaluated each of these criteria in the data. Indeed, the tumor fraction in patches sampled by the RL agent at high resolution (10\% with SASHA-0.1 or 20\% with SASHA-0.2) significantly exceeds that in a corresponding number of randomly chosen patches, among those not sampled by the agent (\cref{fig:figure_3}e, p<0.001). Moreover, the RL policy guides the agent to preferentially samples patches afforded a high attention score by sota models: both the average attention score (\cref{fig:figure_3}g, p<0.001) and the fraction of top-k patches sampled under the RL policy were statistically significantly greater than the fraction of patches sampled by a random policy (\cref{fig:figure_3}f, p<0.001). In other words, our RL agent's sampling actions were guided by an efficient, intuitive and explainable policy.


\vspace{-2mm}
\subsection{Effect of observation budget: Calibration} 
\label{sec:calibration}
\vspace{-2mm}
We have shown that SASHA achieves near-sota performance at a fraction of the inference time, even with sampling 10\% or 20\% of the patches at high resolution. Is there any advantage to sampling more patches, other than achieving a marginal improvement accuracy, especially when the inference time increases almost linearly with the proportion of high-resolution patches sampled. Here, we show empirically that sampling a higher fraction of patches yields better calibrated predictions. With well calibrated models, the model's confidence in its prediction (e.g., predictive uncertainty) is a useful indicator for when referral to a specialist is warranted \citep{Dj_and_jim}. 
We quantified the expected calibration error (ECE) \citep{guo2017calibration} as a measure of the degree of calibration of the models. The ECE decreased systematically at the proportion sampled patches increased from 10\% (SASHA-0.1) to 20\% (SASHA-0.2) to 100\% (HAFED). This pattern occurred possibly because providing the model more timesteps to sample the WSI in each RL episode enabled the model to recover from sub-optimal sampling actions in the early timesteps, and to sample the relevant portions of the WSI in the later timesteps. Surprisingly, ECE with SASHA-0.2 was lower than that of even sota attention-based models like ACMIL and DTFD (\cref{fig:figure_3}h). In other words, while a shorter observation budget encouraged efficient learning, a longer budget produced better calibrated predictions.

\vspace{-2mm}
\section{Conclusions and Limitations}
\label{sec:conc}
\vspace{-3mm}
We show that an attention-augmented deep RL agent outperforms or performs on par with sota models on WSI classification, albeit at a fraction of their memory and computational costs. Key limitations of the model include a significant impact of the observation budget on model calibration, and the need to train the model with all high resolution patches, which significantly increases training time; other limitations are discussed in Appendix \ref*{appendix:limitations}. Nonetheless, we demonstrate explainable,  reliable, and calibrated predictions, which could enhance clinicians' trust in model predictions. 

\vspace{-2mm}
\section*{Acknowledgements}
\label{sec:ack}
\vspace{-3mm}
This research was supported by a Department of Science and Technology SwarnaJayanti Fellowship, a Google Research Grant, a Pratiksha Trust Intramural Grant and a Gore Subraya Bhat Chair Associate Professorship in Digital Health (all to DS), Ministry of Education Scholarships (to TG and NM), an Institute of Eminence (IoE) Grant (to IISc), a Walmart Centre for Tech Excellence at IISc, the Indo–French Centre for the Promotion of Advanced Research Grant (CEFIPRA; Project 7102–1), Kotak IISc FinTech Grant, a DST–SERB Core Research Grant (CRG/2021/008330), and a Pratiksha Trust Young Investigator Award (all to GT). We are very grateful to Karm Patel, Mainak Biswas, Aditi Saxena, and Tej Pratap Yadav for help with model development and simulations, and Raj Jain for help with the original draft of the manuscript. The authors declare no competing interests.




%% file: sec/6_appendix.tex
\begin{center}
    \LARGE \textbf{Sequential Attention-based Sampling for Histopathological Analysis}
\end{center}

\appendix
\section{Appendix}

\subsection{Dataset details}
\label{appendix:dataset}

The CAMELYON16 dataset comprises 399 whole-slide images (WSIs) of lymph node sections. After pre-processing and segmentation, we generated non-overlapping patches of size $256 \times 256$ ($\times 3$ RGB channels). On average, this resulted in, on average, $\sim$800 patches per WSI at a scanning resolution of $5\times$, and $\sim$12{,}500 patches at a high resolution of $20\times$. We used the official test set for evaluation \citep{bejnordi2017diagnostic}. The training set was further divided to create a validation set (10\%) while maintaining a similar proportion of normal and tumor slides in each subset.

The TCGA-NSCLC dataset comprises 1,054 whole-slide images (WSIs) obtained from the TCGA-NSCLC repository under the TCGA-LUSC and TCGA-LUAD projects. Ten low-quality slides without magnification information were discarded, and three slides with errors were ignored. We followed the same test split as in DSMIL \citep{dsmil2021}. The test set comprised a total of 213 slides: 104 from TCGA-LUAD and 109 from TCGA-LUSC. Following preprocessing, we obtained, on average, $\sim$212 patches per WSI at $5\times$ magnification and $\sim$3{,}393 patches at $20\times$ magnification. The training set was further divided to create a validation (20\%) set while maintaining a similar proportion of labels in each subset.

\begin{table}[ht]
    \centering
    \resizebox{\textwidth}{!}{
        \renewcommand{\arraystretch}{1.5}
        \begin{tabular}{llcccccc}
        \toprule
        
        \multirow{2}{*}{\bf Dataset} & \multirow{2}{*}{\bf Task} & \multicolumn{2}{c}{\bf Train} & \multicolumn{2}{c}{\bf Val} & \multicolumn{2}{c}{\bf Test} \\
        \cline{3-8}
        & & \textbf{Class 1} & \textbf{Class 2} & \textbf{Class 1} & \textbf{Class 2} & \textbf{Class 1} & \textbf{Class 2} \\
        
        \midrule
        
        CAMELYON16 & Classify tumor & 140 & 102 & 18 & 9 & 80 & 49 \\
        TCGA-NSCLC & Classify sub-type & 339 & 319 & 86 & 81 & 104 & 109 \\
        
        \bottomrule
        
    \end{tabular}
    }
    \vspace{6pt}
    \caption{Summary of datasets used in this study. CAMELYON16 involves binary classification of normal versus tumor slides, while TCGA-NSCLC involves distinguishing between adenocarcinoma (LUAD) and squamous cell carcinoma (LUSC). The table shows the number of WSIs for each class in the training, validation, and test sets. Here, ``Class 1'' refers to Normal (for CAMELYON16) and LUAD (for TCGA-NSCLC), while ``Class 2'' refers to Tumor (for CAMELYON16) and LUSC (for TCGA-NSCLC).}
    \label{table:dataset_summary}
\end{table}

The CamelyonPlus dataset\citep{ling2025camelyonplus}, comprises 1,347 whole‐slide images (WSIs) derived from the curated CAMELYON16 and CAMELYON17 collections, after removal of low-quality slides and re-labeling. It involves a four-class classification. We created training, validation, and test splits to ensure class balance across all four labels. A summary of the splits and labels can be found in \cref{table:camelyon_plus_summary}.

\begin{table}[ht]
    \centering
    \resizebox{\textwidth}{!}{
        \renewcommand{\arraystretch}{1.5}
        \begin{tabular}{lcccccccccccc}
        \toprule
        
        \multirow{2}{*}{\bf Dataset} &   
        \multicolumn{4}{c}{\bf Train} & \multicolumn{4}{c}{\bf Val} & \multicolumn{4}{c}{\bf Test} \\
        \cline{2-13}
        &  \textbf{Neg.} & \textbf{Mi-m.} & \textbf{Ma-m.} & \textbf{ITC} &
        \textbf{Neg.} & \textbf{Mi-m.} & \textbf{Ma-m.} & \textbf{ITC} &
        \textbf{Neg.} & \textbf{Mi-m.} & \textbf{Ma-m.} & \textbf{ITC} \\
        
        \midrule
        
        Camelyon+ & 
        608 & 122 & 174 & 36 &
        130 & 26 & 38 & 9 &
        131 & 26 & 38 & 9 \\
        
        \bottomrule
        
    \end{tabular}
    }
    \vspace{6pt}
    \caption{Summary of Camelyon+ dataset. The WSIs have the following 4-class slide-level labels: Negative (Neg.), Micro-metastasis (Mi-m.), Macro-metastasis (Ma-m.), and Isolated Tumor Cells (ITC). The table shows the number of WSIs for each class in the training, validation, and test sets.}
    \label{table:camelyon_plus_summary}
\end{table}


The BRACS (BReAst Carcinoma Subtyping) dataset comprises 547 labeled WSIs obtained from the public repository \citep{brancati2022bracs}. This dataset involves a 7-class classification task, characterizing each tumor as one of three benign (BT), two atypical (AT), or two malignant (MT) sub-types. We ignored the ``normal'' subtype in the benign (BT) class, and each set of subtypes within BT, AT, and MT was grouped into its respective class. Additionally, a small number of slides (7/503) that lacked all relevant levels of magnification were excluded. After preprocessing, we use the official dataset split across train, validation, and test split with details outlined in \cref{table:appendix:bracs_dataset}. 

\begin{table}[hb]
\centering

\begin{tabular}{lccc}
\toprule
 & \textbf{Benign (BT)} & \textbf{Atypical (AT)} & \textbf{Malignant (MT)} \\
\midrule
Train & 171 & 52 & 135 \\
Val   & 16  & 14 & 21  \\
Test  & 25  & 23 & 16  \\
\bottomrule
\end{tabular}
\vspace{6pt}
\caption{BRACS dataset distribution for 3-class classification (Benign, Atypical, Malignant) across training, validation, and test splits.}
\label{table:appendix:bracs_dataset}

\end{table}

\subsection{Implementation and Hyper-parameter Selection}
\label{appendix:implen_and_hy-params}

\subsubsection{Patch feature distillation with HAFED}
\label{appendix:hafed}
The first step of SASHA involves feature extraction and aggregation of high-resolution patch features, which specifies the RL agent's state $S(t)$. We use a Vision Transformer (ViT) pretrained on histopathological images to extract features from each zoomed in, high-resolution patch \citep{kang2023ssl}. This ViT was used {\it as is} without any additional fine tuning, and its output -- a 384 dimensional embedding per patch -- was provided as input to a hierarchical attention-based feature aggregator and distiller (HAFED). 

\begin{table}[h!]
\centering
\begin{tabular}{lcccc}
\toprule
\textbf{Layer name} & \textbf{Input Dim} & \textbf{Output Dim} & \textbf{Input Batch} & \textbf{Output Batch} \\

\midrule

Layer-1 & 384 & 128 & $U_i=\{U_{i1}, U_{i2}, \ldots, U_{ik}\}$ & $A_i=\{a_{i1}, a_{i2}, \ldots, a_{ik}\}$ \\
Layer-2 & 384 & 128 & $U_i=\{U_{i1}, U_{i2}, \ldots, U_{ik}\}$ & $B_i=\{b_{i1}, b_{i2}, \ldots, b_{ik}\}$ \\
Layer-3 & 128 & 1   & $\mathrm{sigmoid}(A) \cdot \tanh(B)$ & ${\alpha_1}^i \in \mathbb{R}^k$ \\
\midrule

\multicolumn{5}{c}{$V_i = {\alpha_1}^i \cdot U_i,\quad V_i \in \mathbb{R}^{d}$} \\
\midrule
Layer-4 & 384 & 128 & $V=\{V_{1}, V_{2}, \ldots, V_{N}\}$ & $C=\{c_{1}, c_{2}, \ldots, c_{N}\}$ \\
Layer-5 & 384 & 128 & $V=\{V_{1}, V_{2}, \ldots, V_{N}\}$ & $D=\{d_{1}, d_{2}, \ldots, d_{N}\}$ \\
Layer-6 & 128 & M (5) & $\mathrm{sigmoid}(C) \cdot \tanh(D)$ & ${\alpha_2} \in \mathbb{R}^{N \times M}$ \\
\midrule

\multicolumn{5}{c}{$\alpha = \frac{1}{M} \sum_{j=1}^{M} \alpha_2^{(:,j)}, \quad \alpha \in \mathbb{R}^{N}$; $\quad h = {\alpha} \cdot V, \quad h \in \mathbb{R}^{d}$} \\
\midrule
Layer-7 & 384 & 2 & $h$ & logits \\

\bottomrule

\end{tabular}
\vspace{6pt}
\caption{Architecture of the HAFED model. Layers 1-3 correspond to the feature aggregator (FA) and layers 4-7 correspond to the classifier. $N$ ($\sim$800 for camelyon-16) is the number of patches in low-resolution WSI, $k$ (16) is the number of sub-patches in high-resolution per one patch in low-resolution, $d$ (384) feature embedding dimension of one patch. $U_i\in \mathbb{R}^{16 \times 384}$, $A_i \in \mathbb{R}^{16 \times 128}$, $B_i \in \mathbb{R}^{16 \times 128}$, $C \in \mathbb{R}^{N \times 128}$, $D \in \mathbb{R}^{N \times 128}$, $h \in \mathbb{R}^{384}$.
}
\label{table:hafed_arch}
\end{table}

As described in \Cref{sec:fac}, the first stage of HAFED acts as a feature aggregator (FA); the architecture of this network is shown in \cref{table:hafed_arch}. 
We employed a single attention branch, which takes high resolution features \(U \in \mathbb{R}^{N \times k \times d}\) and assigns attention scores for each of the $k$ sub patches. The inner product between the attention scores and $U$ produces $V \in \mathbb{R}^{N \times d}$, which is in same dimensionality as the low-resolution features ($Z \in \mathbb{R}^{N \times d}$). The second stage of HAFED acts as a classifier, which is trained end-to-end with the feature aggregator. Here, we use five attention branches for the CAMELYON16 dataset and one for the TCGA-NSCLC dataset; the number of branches was determined as a hyperparameter based on the respective validation set (\cref{tabel:hyper-params}). This stage takes $V$ as input, assigns attention scores across the N patches, and computes the inner product between the scores and $V$ to produce a 384-dimensional embedding representing the entire WSI. The architecture of this network is shown in \cref{fig:hafed_tsu}a. We applied masking to the top four and top eight patches with a probability of 0.4 in the first and second stages, respectively (\cref{tabel:hyper-params}).

HAFED is trained with an annealed learning rate schedule starting at 4 \(\times10^{-4}\), with the AdamW optimizer and a weight decay of 1 \(\times10^{-4}\). The HAFED module is trained end-to-end using the following loss components: (1) a cross-entropy loss between the final predicted label and the ground truth label, (2) a cross-entropy loss applied independently to the predictions from each attention branch in second stage and ground truth label, and (3) a dissimilarity loss that encourages diversity among the attention scores of the different branches in the second stage. This dissimilarity loss encourages each branch to focus on distinct tumor features, thereby improving the model’s ability to capture complementary diagnostic features\citep{zhang2023acmil}. 

During inference, at each time step, the first stage of HAFED (FA), receives as input $k$ zoomed-in sub-patch images corresponding to the selected action. It assigns attention scores to estimate the relative importance of each sub-patch and aggregates their features by computing a weighted sum based on these scores; this is then used to update the state (see next subsection, \cref{appenidx:tsu}). In the final time step, the second stage of HAFED (classifier), takes the state representation  $S_T$ and predicts the label $\hat{y}$. In addition to the final classification, the second stage is also used to assign rewards during SASHA' RL agent training (see subsection \cref{appendix:actor_critic}).

\begin{table}[h!]
\centering
\footnotesize

\resizebox{\textwidth}{!}{

\begin{tabular}{lccccccc}

\toprule

\textbf{Model} & \textbf{Hyper-parameters} & \textbf{Selection Criterion} & \multicolumn{4}{c}{\textbf{SASHA - 0.1 / 0.2}} \\ 
 & & & \multicolumn{2}{c}{\textbf{CAMELYON16}} & \multicolumn{2}{c}{\textbf{TCGA-NSCLC}} \\ 
 & & & \textbf{Range} & \textbf{Values} & \textbf{Range} & \textbf{Values} \\

\midrule

\multirow{3}{*}{HAFED} 
  & Attn-2 in HAFED (M) & Validation & 1-5 & 5 & 1-5 & 1 \\
  & Attn -1 Masking Probability & Validation & 0.4 - 0.6   & 0.4 & 0.4 - 0.6 & 0.4 \\
  & Attn -2 Masking Probability & Validation & 0.4 - 0.6   & 0.4 & 0.4 - 0.6 & 0.4 \\
  
\midrule

\multirow{1}{*}{TSU} 
  & Cosine Threshold in TSU $(\tau)$ & Validation & 0.88 - 0.98 & 0.9 / 0.95 & 0.8 - 0.92 & 0.9/0.9 \\
  
\midrule

\multirow{4}{*}{Actor \& Critic} 
  & Discount Factor & Validation & 0.8-1.0 & 1.0 & 0.8-1.0 & 1.0 \\
  & Clipping Parameter & Default & - & 0.1 & - & 0.1 \\
  & Entropy Regularization Coefficient & Default & -& 0.001 & - & 0.001\\
  & Gradient Clipping Threshold & Default & -& 0.5 & - & 0.5 \\
  
\bottomrule

\end{tabular}

}
\vspace{6pt}
\caption{Hyperparameters for the CAMELYON16 and TCGA-NSCLC datasets. In the Values column, entries separated by a slash (/) represent values for SASHA-0.1 and SASHA-0.2 models respectively; a singleton entry reflects the same value across both SASHA variants.
}

\label{tabel:hyper-params}
\end{table}

\subsubsection{Updating WSI state with TSU}
\label{appenidx:tsu}
The second step of SASHA involves updating the RL agent's state with the high-resolution patch representation extracted with HAFED. During training, the states of similar patches are concurrently updated with a targeted state update (TSU) rule. Specifically, this included all patches whose low-resolution representation's similarity to the currently selected patch, exceeded a threshold value ($\tau$), based on cosine similarity; $\tau$ was tuned as a hyperparameter (\cref{tabel:hyper-params})

The architecture of this TSU network is shown in  \cref{table:fglobal_model_summary}. For each patch $a_{\tau}$ that crosses the similarity threshold, the TSU model takes as input a concatenated vector comprising (i) its low-resolution features $Z_{a_{\tau}} \in \mathbb{R}^{d}$, (ii) the low-resolution feature representation of the selected patch $a_t$, $Z_{a_{t}} \in \mathbb{R}^{d}$ and (iii) the high-resolution feature representation of the selected patch $V_{a_{t}} \in \mathbb{R}^{d}$. Note that while (i-ii) are obtained directly from pretrained ViT, (iii) is obtained from the FA output of the HAFED model. Concatenating these 3 features results in a $3d$-dimensional input ($3 \times 384 = 1152)$, based on which the model predicts a $d$-dimensional high-resolution feature of the patch $a_{\tau}$, $\hat{V}_{a_{\tau}}$. The state of the selected patch alone is updated by directly copying its high-resolution HAFED (FA) output $V_{a_{t}}$, into its (corresponding) entry in the WSI state $S(t)$. In this manner, the TSU integrates information from the low-resolution representation of patches similar to the currently visited patch, with the high-resolution representation of visited patches to update the WSI state. 

During training the TSU network is trained by sampling patches randomly, and updating their states based on the predicted high-resolution feature of each patch ($\hat{V}$) and its actual high-resolution feature ($V$) obtained from HAFED's feature aggregator. Once a patch state is updated -- either directly, by visiting it, or indirectly, by the similarity-based rule -- its index is masked out, and it is no longer visited during training. The TSU model is trained for 500 epochs with an Adam optimizer and a learning rate of 1 \(\times10^{-4}\). The best model was chosen based on the minimum of the validation loss.

During inference, the low-resolution patch to visit at each time step $a_t$ is selected by RL agent. The state of patches similar to the currently visited patch are updated with the same sequence of steps as described for the training. A key difference is that, in this case, only the directly visited patch indices are masked out, and the RL agent is free to select, for its action in the next time step ($t+1$), any of the patches that have not been visited directly in the previous time steps (${0, \ldots, t}$).

\begin{table}[h!]
\centering

\begin{tabular}{lccr}
\toprule
{\bf Layer} & {\bf Input Dim} & {\bf Output Dim} & {\bf \# Parameters} \\

\midrule
FC-1 & 1152 & 768 & 885,504 \\
FC-2 & 768 & 384 & 295,296 \\
Layer Norm & - &  384 & 384 \\

\bottomrule

\end{tabular}
\vspace{6pt}
\caption{Architecture of the TSU model. The input is a $3d = 1152$ dimensional vector ($d = 384$; see text for details). The final output is $d$-dimensional. FC: fully connected layer with bias.}
\label{table:fglobal_model_summary}

\end{table}

\subsubsection{Actor and Critic}
\label{appendix:actor_critic}

The final component of SASHA is a reinforcement learning (RL) agent responsible for sampling, sequentially, the most relevant patches for high-resolution analysis. This selection is formulated as a sequential decision-making problem, where the agent learns to identify diagnostically informative regions within each slide. We adopt the Actor-Critic framework and train the agent using the Proximal Policy Optimization (PPO) algorithm, a widely used on-policy method known for its stability and effectiveness in complex RL tasks \citep{schulman2017ppo}.

The purpose of the actor network is to learn a policy -- a conditional probability distribution over actions given the current state $S_t$. In our setup, the environment is described as, the state is initialized with low-resolution features ($Z \in \mathbb{R}^{N \times d}$), the action is selected from a discrete action space $A \in \{1,2, \ldots, N\}$ (indices of low-resolution patches), and the reward  function is specified as the negative of cross-entropy (CE) loss between the predicted WSI label, at time step $t,$ and the ground truth. Given the definitions of state, action and reward, the actor and critic networks are designed as follows. 

The Actor network follows the same architecture as Layers 1–3 in \Cref{table:hafed_arch}. A softmax is applied to the output attention scores to obtain a categorical distribution (i.e., the policy) over the $N$ patches, from which an action $a_t$ (i.e., a patch index) is sampled during training.
To reduce redundancy, once a patch is selected, its index is masked in all subsequent time steps -- by setting its corresponding logit to negative infinity -- effectively removing it from consideration for future visits. the rationale behind this policy is that zooming into the same patch multiple times will not yield additional information, as the patch content remains unchanged.
After selecting $a_t$, the Target State Updater (TSU) module is invoked to update the current state $S_t$ to the next state $S_{t+1}$, thereby incorporating the high-resolution information from the selected patch into the state representation.

The Critic network also mirrors the backbone of actor network i.e architecture of Layers 1–3 in \Cref{table:hafed_arch}, with an additional final layer that takes the aggregated state representation ($\alpha \cdot S_t$), and predicts the reward-to-go or expected return given the current state $S_t$. The reward-to-go is used in calculating the generalized advantage estimate (GAE, \citep{gae}). During training the actual reward-to-go from the given state is known, and the critic network is trained by minimizing MSE loss between the predicted and actual value.

Both networks are trained for 15 epochs using the AdamW optimizer with an annealing learning rate starting from $1 \times 10^{-5}$ and a weight decay of $1 \times 10^{-3}$. Network hyperparameters are reported in \cref{tabel:hyper-params}. During training of the Actor and Critic networks, all other components of SASHA are frozen (i.e., HAFED, TSU), and only the actor and critic networks are updated. We select the best policy across epochs as the one that minimizes the validation loss while simultaneously maximizing the sum of the validation AUROC and F1 scores ($\mathrm{min}_{\pi} [\mathrm{Loss + 2 - (AUROC+F1)}]$). This formulation ensures that the selected policy balances both accuracy (AUROC and F1) and calibration (validation loss).


During inference, given the current state $S_t$, the Actor network outputs a policy in the form of a categorical probability distribution over the $N$ patches. The action $a_t$ is then selected deterministically as the patch index with the highest probability. In a later section (\cref{appendix:alt_policies}) we explore the effect of alternative, stochastic selection policies during inference.

    
    

\subsection{Additional Control Experiments}
\label{appendix:add_ablation}

\subsubsection{Multiclass Classification Performance of SASHA}
\label{appendix:multi_cls_sasha}
We extended the SASHA framework to evaluate its performance on multiclass classification tasks using the Camelyon+ and BRACS datasets (see \cref{appendix:dataset} for a description of the datasets).

For both datasets, the SASHA framework was trained using experimental settings identical to those of the binary classification tasks. The same patch-level sampling strategy and aggregation mechanism were employed to ensure consistency in evaluation. To mitigate class imbalance, we employed a resampling mechanism that oversamples minority classes (e.g., ITC for the Camelyon+ dataset) during training. 

The performance comparison (\cref{tab:multi_cls_performance}) shows that the methods (HAFED, SASHA) perform comparably with other sota methods, even in this multiclass setting. As before, sampling only 20\% of the patches (SASHA-0.2) achieves performance that is comparable to methods that sample all of the WSI patches at high resolution (e.g., DTFD, ACMIL).


\begin{table}[h] 
\centering

\resizebox{\textwidth}{!}{

\begin{tabular}{llcccccc}
\toprule
\multirow{2}{*}{} & \multirow{2}{*}{Method} & \multicolumn{3}{c}{CAMELYON+} & \multicolumn{3}{c}{BRACs} \\
& & Accuracy & AUC & F1 & Accuracy & AUC & F1 \\
\midrule

\multirow{3}{*}{\rotatebox[origin=c]{90}{100\%}} 
& DTFD              & 0.848$\pm$0.022 & 0.921$\pm$0.023 & 0.804$\pm$0.04 & 0.570$\pm$0.035 & 0.782$\pm$0.032 & 0.570$\pm$0.035\\
& ACMIL             & 0.832$\pm$0.013 & 0.898$\pm$0.005 & 0.786$\pm$0.023 & 0.534$\pm$0.046 & 0.749$\pm$0.014 & 0.596$\pm$0.059\\
& {HAFED} (ours)            & 0.866$\pm$0.014 & 0.934$\pm$0.01 & 0.847$\pm$0.015 & 0.544$\pm$0.03 & 0.747$\pm$0.01 & 0.575$\pm$0.036 \\

\midrule
\multirow{1}{*}{\rotatebox[origin=c]{90}{\parbox{0.4cm}{}}}
& SASHA-0.1 (ours)   & 0.869$\pm$0.008 & 0.905$\pm$0.003 & 0.848$\pm$0.013 & 0.596$\pm$0.001 & 0.729$\pm$0.001 & 0.532$\pm$0.010 \\
\midrule
\multirow{1}{*}{\rotatebox[origin=c]{90}{\parbox{0.5cm}{}}}
& SASHA-0.2 (ours)   & 0.871$\pm$0.005 & 0.935$\pm$0.006 & 0.852$\pm$0.006 & 0.598$\pm$0.016 & 0.762$\pm$0.010 & 0.533$\pm$0.079 \\
\bottomrule
\end{tabular}
}
\vspace{6pt}
\caption{Comparison of multiclass classification performance of different methods on the Camelyon+ and BRACS datasets. Reported metrics include weighted-averaged Accuracy, AUC, and F1-score.}
\label{tab:multi_cls_performance}
\vspace{-5mm}
\end{table}

\subsubsection{Alternative Models under an Observation Budget}
\label{appendix:alt_models}
We compared the performance of SASHA with competing sota models like ACMIL \citep{zhang2023acmil} and DTFD \citep{zhang2022dtfd} under the same observation budget of either 10\% or 20\%. That is, we predicted slide labels by having these sota models examine only a limited fraction of patches in each WSI at high resolution, matching the fraction of the WSI sampled by SASHA.

Each model was trained with a combination of low-resolution and high-resolution patch features and tested, similarly, with a combination of high and low resolution patch features in the same proportion as of training. Specifically, for a given slide, if \( N \) is the total number of low-resolution patches and each low-resolution patch corresponds to \( k \) high-resolution patches, then for that slide, the input comprised of a total of \(N_L + N_H \) patches comprising \( N_L = \left(N - \left\lceil f \cdot N \right\rceil \right) \) low-resolution and \( N_H 
 = \left\lceil f \cdot N \right\rceil \cdot k\) high-resolution patches, where \( f \in \{0.1, 0.2\} \). Because low and high-resolution patches were of the same pixel dimensionality $(256, 256, 3)$, the pretrained ViT model (\cref{sec:method}) was used to extract features from each such patch; this yielded a feature vector $W \in \mathbb{R}^{(N_L + N_H) \times d}$ which was provided as input to each model for label prediction, both during training and evaluation. 

We tested these models on the CAMELYON16 dataset. First, we created five random splits of the train and validation sets of the dataset. For each dataset split, we performed random patch sampling with three different random seeds. Then, for each model, we computed the average performance across these 15 (5 splits $\times$ 3 seeds) model combinations. In each case, the sum of validation accuracy and F1 scores was used to determine the model for evaluation.

\begin{table}[h]
    \centering
    \begin{tabular}{lccc}
    \toprule
    
    \textbf{Model} & \textbf{Accuracy} & \textbf{AUC} & \textbf{F1} \\
    
    \midrule
    ACMIL - 0.1 & 0.841$\pm$0.036 & 0.828$\pm$0.034 & 0.773$\pm$0.038 \\ 
    DTFD - 0.1 & 0.829$\pm$0.013  & 0.835$\pm$0.039 & 0.829$\pm$0.013 \\
    SASHA - 0.1  & \textbf{0.901$\pm$0.021} & \textbf{0.918$\pm$0.014} & \textbf{0.856$\pm$0.031} \\

    \midrule
    ACMIL - 0.2 & 0.838$\pm$0.035 & 0.857$\pm$0.030 & 0.773$\pm$0.043 \\
    DTFD - 0.2  & 0.862$\pm$0.027 & 0.878$\pm$0.030  & 0.862$\pm$0.027 \\
    SASHA - 0.2 & \textbf{0.953$\pm$0.017} & \textbf{0.979$\pm$0.008} & \textbf{0.937$\pm$0.024} \\
    
    \bottomrule
    \end{tabular}
    \vspace{6pt}
    \caption{Performance of ACMIL (rows 1, 4), DTFD (rows 2, 5) and SASHA (rows 3, 6) models under a limited observation budget of 10\% (-0.1) or 20\% (-0.2) for high-resolution patches per WSI (see text for details). Values represent averages and standard deviations on the CAMELYON16 dataset across 15 model combinations (5 random data splits $\times$ 3 random seeds for sampling per split).}
    \label{tabel:appendix_ablation_1}
\end{table}

As shown in \cref{tabel:appendix_ablation_1}, such a strategy of sparsely sampling patches substantially reduced the performance of ACMIL and DTFD models. In each case, SASHA sampling either 10\% (0.1 fraction) or 20\% (0.2 fraction) of the patches at high resolution outperformed the sota models with the corresponding, identical observation budget. In other words, competing sota models failed to live up to SASHA's performance under a comparable observation budget for high-resolution patches. 

\subsubsection{Alternative Action Policies during Inference}
\label{appendix:alt_policies}
In the current SASHA implementation, the inference strategy follows a deterministic, greedy policy wherein, for a given state, the action with the highest probability is always selected. This means the agent always picks the most likely action as per the learned policy distribution during inference (but not during training, see \cref{appendix:actor_critic}). 

We evaluate the effectiveness of this policy over other stochastic policies, similar to the ones used during learning. Specifically, we introduce more exploration by considering stochastic policy variation, where the agent samples actions from the entire probability distribution over actions given a state. 

We explore two different variants -- top-$k$ sampling and top-$p$ sampling. In top-$k$ sampling, where only the top $k$ actions with the highest probabilities are retained as candidates for stochastic selection, and the remaining actions are are masked out (assigned zero probability). The agent then samples from this limited distribution, following renormalization. In top-$p$ sampling, actions are sorted in descending order of their probability, and we identify the smallest set of actions such that their cumulative probability exceeds a predefined threshold $p$. Sampling is then done from this subset, ensuring a balance between diversity and confidence in action selection. 

Comparison of different sampling strategies is shown in \cref{table:appendix:different_sample_strategies}. Interestingly, the default deterministic policy of SASHA --  a greedy strategy of selecting the action with the highest probability -- outperformed all other stochastic policies. This behavior was observed empirically, where the agent consistently selects action with highest probability given current state $S_t$, resulting in more informative state transitions from $S_t$ to $S_{t+1}$ and improved classification performance, as shown in \cref{table:appendix:different_sample_strategies}
. In addition, the stochastic policy that had the greatest degree of freedom for sampling (Stochastic - full distribution) outperformed, by and large, the other policies (top-$k$ and top-$p$); this was understandable because the latter policies impose more constraints on the action space for sampling.

\begin{table}[H]
\centering
\normalsize

\begin{tabular}{@{}lccc@{}}
\toprule
\textbf{Inference Policies}        & \textbf{Accuracy} & \textbf{AUC} & \textbf{F1} \\ 

\midrule
Stochastic - full distribution         & \underline{0.884}      & 0.846        & \underline{0.831}     \\ 
Stochastic - top-$k$ ($k=3)$           & 0.876      & 0.826        & 0.814    \\
Stochastic - top-$k$ ($k=5)$           & 0.868      & 0.823        & 0.805    \\
Stochastic - top-$p$ ($p=0.9)$         & 0.876      & \underline{0.859}        & 0.818     \\
Stochastic - top-$p$ ($p=0.8)$         & 0.868      & 0.851        & 0.795     \\

Deterministic (SASHA default)                         & {\bf 0.953}      & {\bf 0.979}        & {\bf 0.937}    \\
\bottomrule
\end{tabular}
\vspace{6pt}

\caption{Comparison of different stochastic policies (top 5 rows), with SASHA's default deterministic policy, during inference. Values denote the performance of the SASHA-0.2 model on the CAMELYON16 dataset. \textbf{Bold} and \textul{underlined}: best and second best performing policies, respectively. }

\label{table:appendix:different_sample_strategies}
\end{table}

\subsubsection{Alternative Reward Schedules during Training}
\label{appendix:alt_rewards}
The default SASHA model employs a step-wise reward during training, where the agent receives a reward at every time step of the episode associated with the label prediction loss at that time step (see \cref{appendix:actor_critic}). In contrast, a previous competing model (RLogist \citep{zhao2022rlogist}) employed a terminal reward strategy, providing feedback only at the end of each episode. 

We tested this alternative reward schedule employing only a terminal reward at the end of each episode, both during training and inference. As shown in \cref{table:appendix:reward_strategies}, the RL agent trained with reward at each time step significantly outperformed an agent trained using terminal rewards. We propose that this occurs because our reward formulation -- which is designed to minimize the classification loss at each time step -- encourages the model to sample more tumor-relevant patches during inference. Confirming this prediction SASHA's default reward schedule sampled, on average, more patches with a high attention (relevance) score, as compared to the terminal reward schedule (\cref{table:appendix:reward_strategies}).  

\begin{table}[h!]
\centering
\normalsize
\resizebox{\textwidth}{!}{
\begin{tabular}{@{}lccccc}
\toprule

\textbf{Reward Schedule} & \textbf{Accuracy} & \textbf{AUC} & \textbf{F1} & \makecell{\textbf{Top-k attention score}\\ \textbf{patches $\cap$ RL actions}} & \makecell{\textbf{Avg attention}\\ \textbf{score of RL actions}} \\

\midrule

Last time step (Terminal)       & 0.892 & 0.941 & 0.862 & 15.5 & 0.0020 \\
Each time step (SASHA default)  & \textbf{0.953} & \textbf{0.979} & \textbf{0.937} & {\bf 24.25} & {\bf 0.0032} \\

\bottomrule
\end{tabular}
}
\vspace{6pt}
\caption{Comparison of a terminal reward schedule (row 1) with SASHA's default schedule of reward at each time step (row 2), during training (see text for details). Column 5 tells the average patches per slide that are selected by RL agent are in top-k according to the attention score given by HAFED model. Column 6 tells the average attention score of the selected patches by RL agent. (see \cref{fig:figure_3}f-g for details). Values denote the performance of the SASHA-0.2 model on the CAMELYON16 dataset. \textbf{Bold}: the best performing reward schedule.}

\label{table:appendix:reward_strategies}
\end{table}

\subsubsection{Analysis of RL Agent Behavior}

\begin{figure}[htbp]
    \centering
   \includegraphics[width=\textwidth]{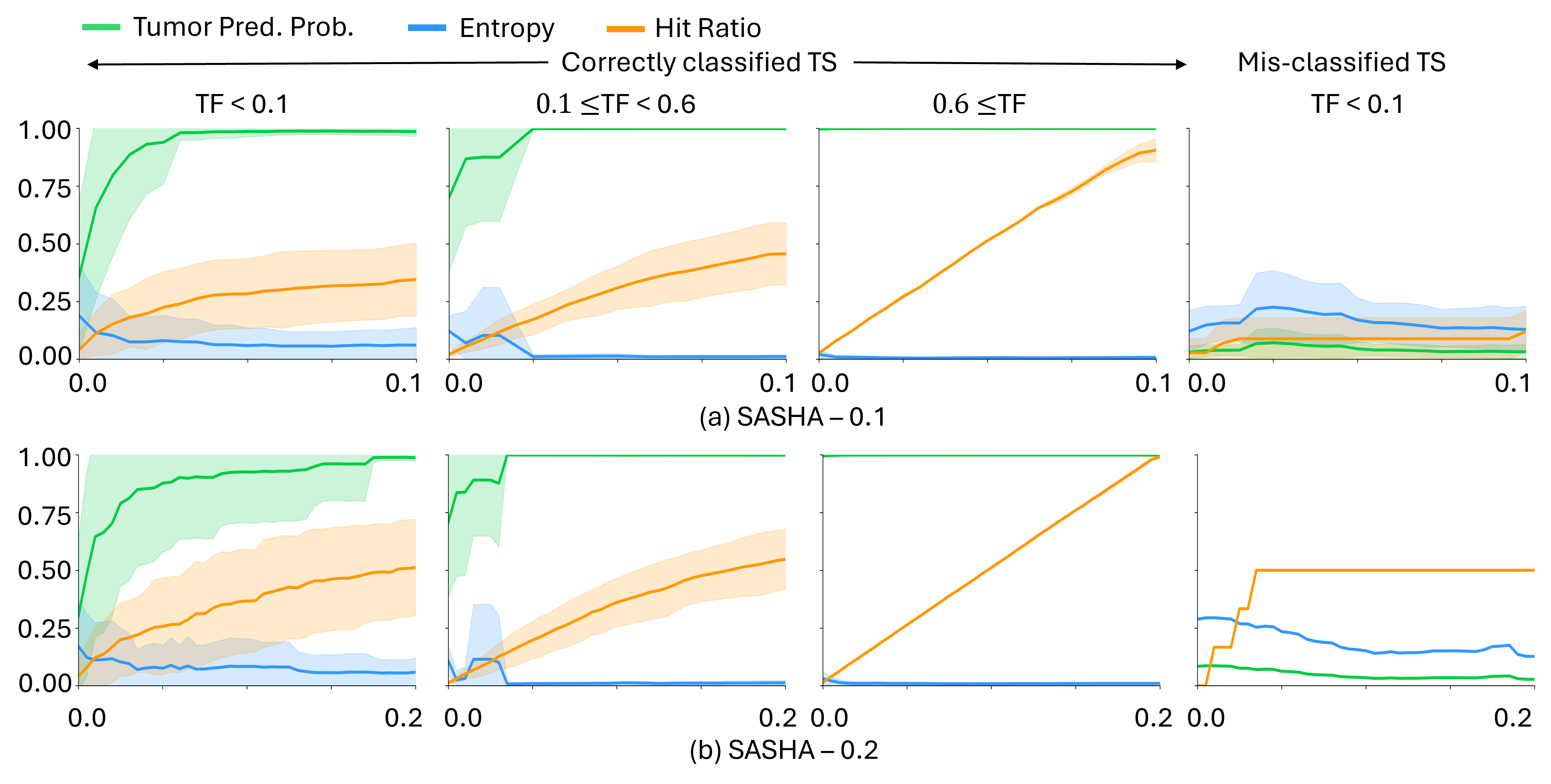}
    
    \caption{
        Evaluation of model performance on tumor slides from the test set, categorized into four groups. The first three columns correspond to correctly classified tumor slides (TS) by both SASHA and HAFED across increasing tumor fraction (TF) ranges, while the fourth column represents TS misclassified by SASHA but correctly predicted by HAFED. The x-axis denotes the visited patch fraction, binned with a width of 0.005. For each slide, we compute the average entropy, average tumor class prediction probability, and maximum cumulative tumor patch hit ratio (\cref{eq:hit_ratio}) within each bin. These metrics are then averaged across all slides, with solid lines indicating the mean and shaded regions representing the standard deviation.
    }
    \label{fig:ap_tsvp}
\end{figure}

To better understand the behavior of the RL agent, we perform two additional analyses. 

First, we asked: how quickly does the RL agent visit tumor patches, and how does the model's confidence evolve over time steps in an episode? We first performed this analysis for slides that were correctly classified by SASHA, separately for 3 classes of WSIs with varying tumor fractions (TF, \cref{fig:ap_tsvp}, columns 1-3). For this, we defined a cumulative tumor patch Hit Ratio using
\begin{equation}
\text{Hit Ratio}(t) = 
\frac{\text{Number of tumor patches sampled up to time} \  t}
{\min\left(\text{Observation Budget},\ \text{Total Tumor Patches in WSI} \right)}.
\label{eq:hit_ratio}
\end{equation}
In other words, the cumulative hit ratio begins at 0.0 at the start of each episode and would linearly reach 1.0 at the end of each episode for the most efficient sampling strategy, i.e. if every patch sampled by the agent were a tumor patch.

For both variants of the model (SASHA-0.1 and SASHA-0.2), on the slides with very sparse tumor regions (TF<0.1), the classifier's output favoring the tumor class reaches a high value ($\sim$0.8-1.0) even before it vists 5\% of all patches at high-resolution (\cref{fig:ap_tsvp}, column 1, green trace). In addition, its predictive entropy drops progressively as more patches are sampled (\cref{fig:ap_tsvp}, column 1, green trace). Surprisingly, even though the peak of the classifier output is reached earlier for SASHA-0.1 as compared to the SASHA-0.2 model, a longer observation budget produces higher accuracies, as well as better calibrated predictions, as shown in the main text \cref{tab:classification_performance} and \cref{fig:figure_3}-h. 

As the budget increases from 10\% to 20\%, the tumor hit ratio improves correspondingly. With a 10\% budget, SASHA tends to misclassify slides with small tumor regions (see F1 score in \Cref{tab:classification_performance}), highlighting the importance of increasing the sampling budget. At 20\%, SASHA correctly classifies nearly all slides that were correctly predicted by HAFED, demonstrating improved robustness across varying tumor fractions.

As the fraction of tumor patches increases (0.1<TF<0.6 and TF>0.6), the model output favoring the tumor class reaches peak upon visiting progressively miniscule fractions of high-resolution patches (<0.025\% or 0.01\%, \cref{fig:ap_tsvp}, columns 2-3, green traces). The predictive entropy also goes to near zero at these fractions indicating confident model predictions with very few samples. 

For \cref{fig:ap_tsvp}, column 2—where the tumor fraction (TF) lies in the range $0.1 \leq \text{TF} < 0.6$—the tumor hit ratio increases slightly as the budget increases. However, the improvement is minimal because most tumor patches are already captured within the initial 10\% of patches selected by the RL agent. This suggests that for WSIs with moderate tumor presence, visiting just 10\% of the patches is often sufficient.


Second, we asked: Is the agent able to localize tumor regions even in slides that are ultimately misclassified? Answering this question is important, because robust tumor localization ability could potentially assist histopathologists by prioritizing regions for first-pass analysis in clinical or laboratory settings. For this analysis, we considered slide that were misclassified by SASHA but correctly classified by HAFED. 

\begin{figure}[htbp]
    \centering

    \resizebox{\textwidth}{!}{%
        \includegraphics{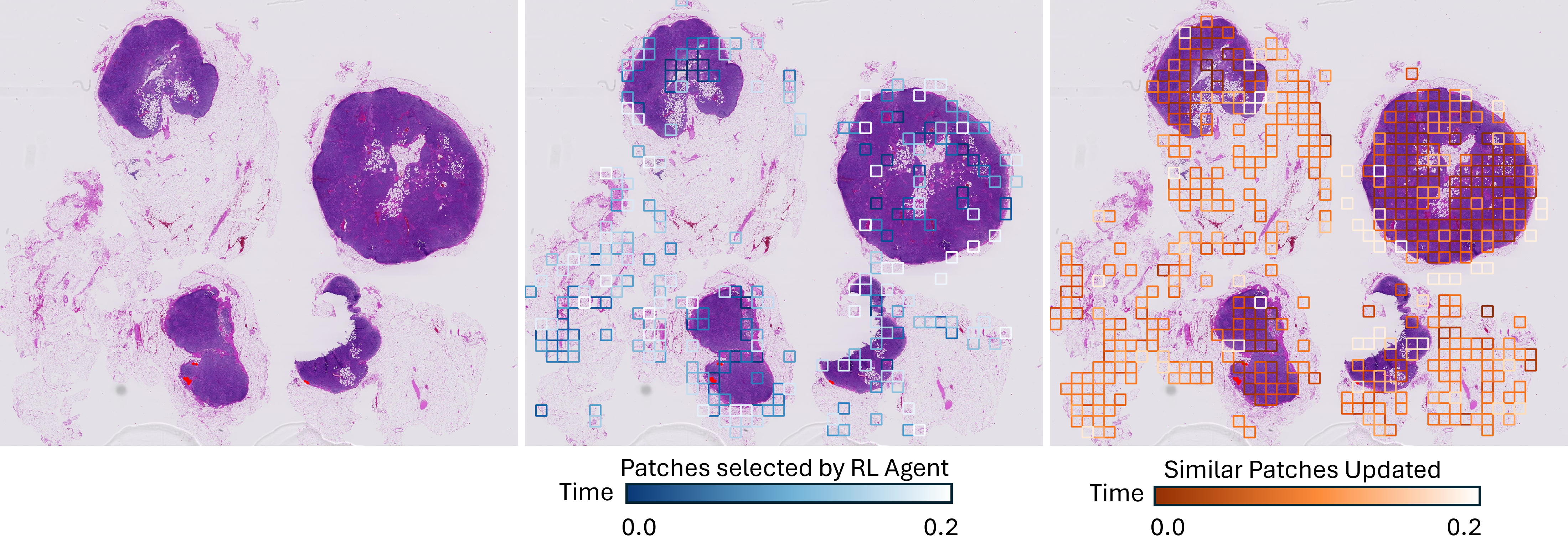}
    }
    
    \caption{Visualization of the observation path traced by the RL agent for a WSI. The first image shows the tumor-annotated region in the WSI. The second image illustrates the patches selected by the RL agent from the initial to the terminal time steps. The third image presents the similar patches updated by the TSU model.
    }
    \label{fig:appendix_visulaization}
\end{figure}

We illustrate this with an example slide \cref{fig:appendix_visulaization}. For this slide, the RL agent successfully identifies tumor patches, with the first tumor patch being visited at a fraction of $0.0085$ and the last at $0.03$. For the entire cohort of such slides, tumor patch visits ranged from $0.001-0.099$ fraction for SASHA-0.1 and $0.009-0.034$ fraction for SASHA-0.2  (\cref{fig:ap_tsvp}, column 4, orange trace). In contrast, when patches are randomly sampled for the same slide, the first tumor patch is typically sampled after visiting $0.141 \pm 0.121$ fraction of the patches. This demonstrates that the RL agent is able to localize tumor regions, typically with an order of magnitude fewer steps than random sampling. 



\begin{table}[htbp]
\centering
\begin{tabular}{lccccccc}
\toprule
\textbf{Method} & \textbf{Patch} & \textbf{FE(HR)} & \textbf{FE(LR)} & \textbf{Infer} & \textbf{TSU} & \textbf{RL} & \textbf{Total} \\
\midrule
DTFD      & 1.46 & 116.75 & N/A   & 0.0230 & N/A  & N/A   & 116.77 \\
ACMIL     & 1.46 & 116.75 & N/A   & 0.0140 & N/A  & N/A   & 116.76 \\
HAFED     & 1.46 & 116.75 & N/A   & 0.0010 & N/A  & N/A   & 116.75 \\
SASHA-0.1 & 1.46 &   8.60 & 4.74  & 0.0007 & 2.06 & 0.105 & 15.50  \\
SASHA-0.2 & 1.46 &  18.23 & 4.95  & 0.0007 & 4.09 & 0.219 & 27.49  \\
\bottomrule
\end{tabular}
\vspace{6pt}
\caption {Comparison of inference time for different methods, including the time involved in each computational step. ``Patch'' denotes the time for dividing the slide into patches, FE denotes Feature Extraction time, with HR representing High Resolution and LR representing Low Resolution. ``Infer'' refers to the inference time for a single model pass. TSU corresponds to the time required for state update, and RL indicates the time needed for agent policy and value updates. The “Total” time (last column) corresponds to the values shown in \cref{fig:figure_3}c, which does not include the fixed “Patch” time overhead. For SASHA, the value in “Infer.” corresponds specifically to the time for one forward pass through the HAFED model.}
\label{table:appendix:inference_time_detail_comparsion}
\end{table}

\subsection{ Computation of WSI Compressibility, Inference time, and Calibration Error}
\label{appendix:comp_of}

Whole Slide Images (WSIs) are typically acquired at gigapixel resolution. For laboratories or hospitals maintaining long-term archives, this poses a significant storage challenge. However, we propose that, for accurate tumor diagnosis, it is not necessary to store WSIs as raw images. By storing them as informative feature representations, a considerable reduction in storage requirements can be achieved.

To quantify the data reduction achieved during WSI processing, we define the \textit{WSI compressibility factor} as the ratio between the original image size at full magnification and the size of the final feature representation used for classification. We define the raw image size as $W \times H \times 3$, representing the width, height, and RGB color channels, respectively.

Models such as ACMIL, DTFD, and HAFED operate on high-resolution feature tensors $U \in \mathbb{R}^{N \times k \times d}$, where $N$ is the number of low-resolution patches, $k$ (16) is the number of sub-patches in high-resolution per low-resolution patch, and $d$ (384) is the feature dimension. In contrast, SASHA operates on an initial low-resolution feature set $Z \in \mathbb{R}^{N \times d}$ and iteratively refines it by zooming into diagnostically relevant patches. The final state representation $S_T \in \mathbb{R}^{N \times d}$ can be archived as a compact summary of the WSI, requiring approximately 16 times less storage compared to methods like ACMIL, DTFD, or HAFED, \cref{fig:figure_3}d.

The compressibility factor $C$ is defined as:
\[
C = \frac{\text{Image Size}}{\text{Feature Space}} =
\begin{cases}
\frac{W \times H \times 3}{N \times k \times d} & \text{for ACMIL, DTFD, HAFED} \\
\frac{W \times H \times 3}{N \times d} & \text{for SASHA}
\end{cases}
\]

Empirically, the compressibility factor is $\sim 5\times10^4$ for ACMIL, DTFD, and HAFED, and $\sim 10^6$ for SASHA-0.1 and SASHA-0.2. This indicates that SASHA methods yield significantly more compact feature representations for classification tasks.

{\bf Inference time computation}: As shown in \Cref{fig:figure_3}c, SASHA, with its selective sampling strategy, achieves the lowest inference time compared to models such as ACMIL, DTFD, and HAFED, which sample all the patches at high resolution. We report the average inference time per slide for the CAMELYON16 test dataset. The inference time reported in \cref{fig:figure_3}c includes all steps: pre-processing, feature extraction, and classification.  \cref{table:appendix:inference_time_detail_comparsion} shows the inference time involved in each step. The column “Patch” refers to the time required to divide the WSI at low resolution into non-overlapping 256$\times$256 patches, while also removing background content such as regions with only the glass slide or insufficient biological tissue. This step is a constant overhead across all methods and is not included in the Total time in \cref{table:appendix:inference_time_detail_comparsion}. Feature extraction of high-resolution patches reflects a major inference time bottleneck. Because SASHA visits only a small fraction (10-20\%) of patches at high resolution, this yields considerable savings over other approaches that visit all (100\%) of patches at high resolution, like DTFD, ACMIL, HAFED. In addition to feature extraction, the TSU is a secondary bottleneck in SASHA because at each timestep, the cosine similarity with all other non-masked patches must be computed, albeit at low resolution. Improving the speed of this step is scope for future work. 


{\bf Expected Calibration Error (ECE):} Expected Calibration Error (ECE) is a metric used to quantify how well a neural network's predicted probabilities align with the true probabilities (or observed frequencies) of its predictions~\citep{guo2017calibration}. It measures how reliable a model's confidence levels are in its predictions. A perfectly calibrated model would have an ECE of 0, indicating that the model's confidence scores accurately reflect the true likelihood of its predictions being correct. 

To compute ECE, we first divide the predicted confidence scores (i.e., the maximum softmax probabilities) into \( M = 10 \) equally spaced bins. For each bin, we calculate two quantities: i) Accuracy: the proportion of correctly classified samples in the bin; and ii) Confidence: the average of the predicted probabilities in the bin. These metrics are then used to compute the calibration error as the absolute difference between its accuracy and confidence for each bin, summed across bins~\citep{guo2017calibration}. 




 
HAFED achieves the lowest Expected Calibration Error (ECE) of approximately 0.015, indicating highly calibrated predictions. In contrast, ACMIL and SASHA-0.1 exhibit the highest ECE values, around 0.06. SASHA-0.2 achieves an intermediate ECE of around 0.03, which is still lower than that of ACMIL and DTFD. This suggests that SASHA-0.2 not only matches HAFED in terms of classification accuracy but also performs better than other sota methods in terms of calibrated predictions.

\subsection{Hardware}
\label{appendix:hardware}

For feature extraction and model inference, we utilized an NVIDIA GeForce GTX 1080 Ti GPU with 12\,GB of memory. Feature extraction was carried out with a batch size of 512. Training for HAFED and TSU was also performed on the same GPU. To accelerate the training of the RL agent described in \cref{sec: sasha}, we employed an NVIDIA Tesla V-100 GPU with 32\,GB of memory. 


\subsection{Limitations and Extensions}
\label{appendix:limitations}
{\bf Training time cost} - A key limitation of our method is that, despite being RL-based, it performs feature extraction on all high-resolution patches during training.
To reduce training time, future improvements would involve developing an end-to-end pipeline that trains the RL agent concurrently with the HAFED and the TSU modules.
In such a model, the policy network decides the specific low-resolution patch to sample in real-time, followed by HAFED feature distillation for the zoomed-in patch, and TSU state update for the entire slide, with all steps being trained end-to-end with reward at each time step.
This would be followed by the additional steps of training the policy and value networks, and so on, until the end of each episode, for several episodes. This mimics how RL agents learn, conventionally, in novel environments with sporadic rewards.
With this method, there would be no need to sample every patch at high resolution during training. Yet, key challenges -- with stable,  training of multiple modules concurrently -- may need to be addressed.

{\bf Fixed sampling budget} - In our current setup, we employ a fixed sampling budget per slide (e.g., 10\% or 20\%). However, as shown in \Cref{fig:ap_tsvp}, we observe that for certain slides, the classifier can become confident in its prediction fairly early in the sampling process, rendering further sampling unnecessary. This suggests that a dynamic budget allocation strategy -- guided by the classifier's confidence -- could further reduce inference time without compromising performance.

{\bf Architecture overhead and parameter scaling} - We compare the total number of model parameters—excluding the ViT-based feature extractor (see \Cref{appendix:hafed})—across different methods. DTFD, ACMIL, and HAFED, all of which operate on high-resolution inputs, are relatively lightweight with approximately $82.56\text{K}$, $84.37\text{K}$, and $300\text{K}$ parameters, respectively. In contrast, the SASHA pipeline consists of three distinct modules: HAFED, the Target State Updater (TSU), and the Reinforcement Learning (RL) agent, with parameter counts of $300\text{K}$, $1.1\text{M}$, and $296\text{K}$, respectively—summing to approximately $1.7\text{M}$ parameters. This represents a $3.5\times$ increase in parameters compared to HAFED and a $21\times$ increase relative to DTFD and ACMIL.

{\bf Filtering normal patches} - WSIs typically contain far more normal patches, as compared to tumor patches. This imbalance becomes particularly problematic when the tumor occupies only a small fraction of the slide. In such cases, the RL agent is more likely to encounter and select normal patches -- especially in the early steps -- leading to inefficient exploration and increased risk of a miss. The sparse distribution of tumor regions makes it difficult for the agent to consistently focus on diagnostically relevant areas within a limited sampling budget. A more efficient strategy would involve the early filtering of normal, redundant, or uninformative patches, allowing the agent to concentrate its computational efforts on a smaller, more informative subset of patches. This could improve both performance and inference efficiency.

{\bf Incorporation into clinical workflows} - Our method is particularly suited for deployment in clinical settings, from two key standpoints -- Explainability and Calibration. First, clinicians are likely to trust decisions made by AI agents only if these decisions can be readily and rationally explained. Our RL agent consistently selects high-attention, tumor-rich patches, as compared to normal patches. Moreover, this selection occurs far more frequently than that of a random policy. These selection strategies make the decisions of the RL agent intuitive and easy to explain to clinicians (\Cref{sec:patch_sel_strategy}). 

Second, clinicians are also likely to benefit from a reliable ``confidence'' score from the AI model, to know when they can, and cannot, trust its predictions. We showed that the expected calibration error (ECE) for SASHA=0.2 were consistently lower than those of competing sota attention-based models like ACMIL and DTFD (\Cref{sec:calibration}).
With well-calibrated models (lower ECE), the model’s confidence in its predictions (e.g., predictive uncertainty) is reliable. In this case, it becomes a useful indicator for when referral to a clinician or expert is warranted. In other words, by quantifying predictive uncertainty of the SASHA-0.2 classifier with established approaches (e.g., predictive entropy), our RL model could provide clinicians with valuable additional information about when its prediction can (or cannot) be trusted. Finally, inference time accuracy may depend on the extent of domain shift between the test and training clinical samples; developing approaches that combine domain adaptation with efficient sampling may be relevant to address this challenge.